\documentclass[12pt,preprint]{aastex}

\usepackage{natbib}
\bibpunct{(}{)}{;}{a}{}{,}

\usepackage[utf8]{inputenc}
\usepackage{amsmath}
\usepackage{color}

\usepackage{textcomp}
\usepackage{gensymb}

\usepackage{graphicx}
\usepackage{epstopdf}

\usepackage{hyperref}
\hypersetup{colorlinks,citecolor=blue,linkcolor=blue}

\shorttitle{Evidence of active MHD instability in EULAG-MHD simulations of solar convection}
\shortauthors{Lawson, Strugarek \& Charbonneau}

\newcommand{\ch}[1]{{{#1}}}
\newcommand{\cch}[1]{{{#1}}}

\begin{document}

\title{Evidence of active MHD instability in EULAG-MHD simulations of solar convection}
\author{Nicolas Lawson, Antoine Strugarek, and Paul Charbonneau}
\affil{D\'epartement de Physique, Universit\'e de Montr\'eal, C.P. 6128 Succ.
Centre-ville, Montr\'eal, Qc H3C 3J7, Canada}
\email{nicolas.laws@gmail.ca}
\email{strugarek@astro.umontreal.ca}
\email{paulchar@astro.umontreal.ca}

\begin{abstract}

We investigate the possible development of magnetohydrodynamical instabilities in the EULAG-MHD ``millenium simulation'' of Passos \& Charbonneau (2014). This simulation sustains a large-scale magnetic cycle characterized by solar-like polarity reversals taking place on a regular multidecadal cadence, and in which zonally-oriented bands of strong magnetic field accumulate below the convective layers, in response to turbulent pumping from above in successive magnetic half-cycles. Key aspects of this simulation include low numerical dissipation and a strongly subadiabatic fluid layer underlying the convectively unstable layers corresponding to the modeled solar convection zone. These properties are conducive to the growth and development of two-dimensional instabilities otherwise suppressed by stronger dissipation. 
We find evidence for the action of a non-axisymmetric magnetoshear instability operating in the upper portions of the stably stratified fluid layers. We also investigate the possibility that the Tayler instability may be contributing to the destabilization of the large-scale axisymmetric magnetic component at high latitudes. On the basis of our analyses, we propose a global dynamo scenario whereby the magnetic cycle is driven primarily by turbulent dynamo action in the convecting layers, but MHD instabilities accelerate the dissipation of the magnetic field pumped down into the overshoot and stable layers, thus perhaps significantly influencing the magnetic cycle period. 
Support for this scenario is found in the distinct global dynamo behaviors observed in an otherwise identical EULAG-MHD simulations, using a different degree of subadiabaticity in the stable fluid layers underlying the convection zone.

\end{abstract}

\section{Introduction}

The dynamo-based model of the solar cycle put forth by \citet{Parker:1955km,Parker:1955bb} 
over half a century ago has stood the test of time remarkably well. The joint inductive action of differential rotation and helical turbulence remains at the heart of many contemporary solar cycle models, but helioseismic inversion of the sun's internal differential rotation has brought increased attention to the tachocline, a rotational shear layer straddling the base of the solar convective envelope \citep{Howe:2009aa}, as 
the locus of toroidal field amplification and storage, prior to its buoyant destabilization and rise to the photosphere to produce bipolar active regions \citep{Fan:2009ww}. 
In these models, turbulent induction within the convection zone and/or generation of a surface dipole moment through the decay of active regions provide the regenerative mechanism required to close the dynamo loop \citep[see][for a survey of these various models]{2010LRSP....7....3C}. 
Moreover, various physical mechanisms have been identified, which could power a dynamo contained entirely within the tachocline. For example, \citet{1989A&A...223..343S} \citep[see also][]{2000A&A...359..364O,2000A&A...359.1205O} 
have argued that helical waves growing along toroidal flux tubes stored within the upper stably stratified portion of the tachocline could provide an azimuthal electromotive force able to regenerate the poloidal component in situ;
likewise, \citet{2001ApJ...551..536D} 
have shown that in the presence of rotation, the joint
magnetohydrodynamical (hereafter MHD) instability investigated by \citet{Gilman:1997cr},  
operating in the tachocline, develops a net hemispheric helicity that could provide an analog of the turbulent electromotive force proposed by \citet{Parker:1955bb} 
and mathematically formalized by \citet{Steenbeck:1969aa} 
as the ``$\alpha$-effect''
of mean-field electrodynamics \ch{\citep[see][for an example of flux-transport dynamo models with an $\alpha$-effect originating from such instability]{2001ApJ...559..428D}}.

The vast majority of the solar cycle models built using these various regenerative magnetic field mechanisms operate in the so-called kinematic approximation, whereby the magnetic back reaction on the inductive flows is altogether neglected or incorporated in the models through largely ad hoc parameterizations. Global MHD simulations of solar convection do not suffer from this shortcoming, but it is only recently that advances in computing power and algorithmic design have jointly led to global simulations producing magnetic fields well-organized on global scales as well as undergoing (more or less) regular polarity reversals \citep[e.g.][]{Ghizaru:2010im,Brown:2011fm,Racine:2011gh,Kapyla:2012dg,2013ApJ...778...11M,Nelson:2013fa,2014ApJ...789...35F,Passos:2014kx,2015ApJ...809..149A}. 
However, few of these simulations include a stably stratified fluid layer underlying the convection zone, and those which do often use strongly enhanced dissipative coefficients to ensure numerical stability, which leads to dissipative dynamics in the convectively stable layers.

The EULAG-MHD simulations reported upon in \citet{Ghizaru:2010im} \citep{Racine:2011gh,Passos:2014kx} 
offer an interesting exception. In these simulations numerical stability is enforced via the advection algorithm itself, which effectively provides an adaptive subgrid model introducing only the minimal amount of dissipation required to maintain stability in regions of strong shear in the flow or magnetic field, and very little in smooth regions \citep[see, e.g.,][]{2003PhFl...15.3890D}. 
Such simulations thus offer a unique opportunity to investigate dynamical effects taking place in the stably stratified layers, in particular the occurrence of instabilities otherwise suppressed by strong dissipation. This is the primary aim of this paper. Working with the EULAG-MHD ``millenium simulation'' presented in \citet{Passos:2014kx} 
and briefly described in \S \ref{sec:structure}, we first investigate in \S \ref{sec:magStable} the mechanisms leading to magnetic field accumulation in the stable layers of the simulation. 
In \S \ref{sec:instabEvidence}, following \citet{Miesch:2007fp} 
we then seek evidence for the development of a magnetoshear instability in the stable layer of the simulation and in \S \ref{sec:tayler} extend our analysis to the Tayler instability. We close in \S \ref{sec:scenario} by speculating on the role such instabilities may play in the large-scale magnetic cycle developing in the simulation.

\section{Simulation characteristics}
\label{sec:structure}

The foregoing analyses are based on the EULAG-MHD ``millenium simulation'' presented and analyzed by \citet{Passos:2014kx}. 
This simulation is based on the numerical solution of the anelastic magnetohydrodynamical equations in a thick, gravitationally stratified shell of electrically conducting fluid, rotating at the solar rate, and subjected to thermal forcing driving convection \citep[see][]{Ghizaru:2010im,Racine:2011gh}.
The simulation is performed using EULAG-MHD \citep{Smolarkiewicz:2013hq}, 
a MHD generalization of the robust multi-scale geophysical flow solver EULAG \citep{2003JCoPh.190..601P,Prusa:2008df}. 
We operate EULAG-MHD in its so-called Implicit Large-Eddy Simulation mode, whereby the dissipation required to maintain numerical stability is delegated to the underlying advection scheme, which in this case is analogous to an adaptive subgrid model where the minimal level of dissipation required to maintain stability is introduced only where and when it is required. This allows to reach turbulent regimes on relatively small spatial meshes, in turn allowing temporally extended integrations. This is particularly important in the solar cycle context, considering the vast disparity of timescale between the convective turnover time (hours to days in the outer reaches of our solution domain), and the large-scale magnetic cycle, with its multi-decadal period.

The millenium simulation used in what follows spans 1600 years, 
in the course of which 39 polarity reversals take place. The cycles are quite regular, with a mean period $40.5\pm 1.5\,$yrs, and are well-synchronized across hemispheres. The solution domain spans $0.604\leq r/R\leq 0.96$, discretized on a modest spatial mesh $128\times 64\times 48$ in longitude, latitude and radius. The background stratification is convectively stable below $r/R=0.711$, and mildly unstable above. We use conventional boundary condition on the flow at the lower and upper boundaries, namely impenetrable and stress-free, with an additional friction term introduced at the very base of the stable layer \citep[see, e.g.][]{Alvan:2015gs}, allowing the damping of gravity waves that would otherwise be generated in the stable layer, and which could not be properly resolved on our spatial mesh and thus lead to numerical divergence.

\subsection{The magnetic field and its cycle}
\label{sec:magnetic}

\begin{figure}[hp]
\includegraphics[width=\linewidth]{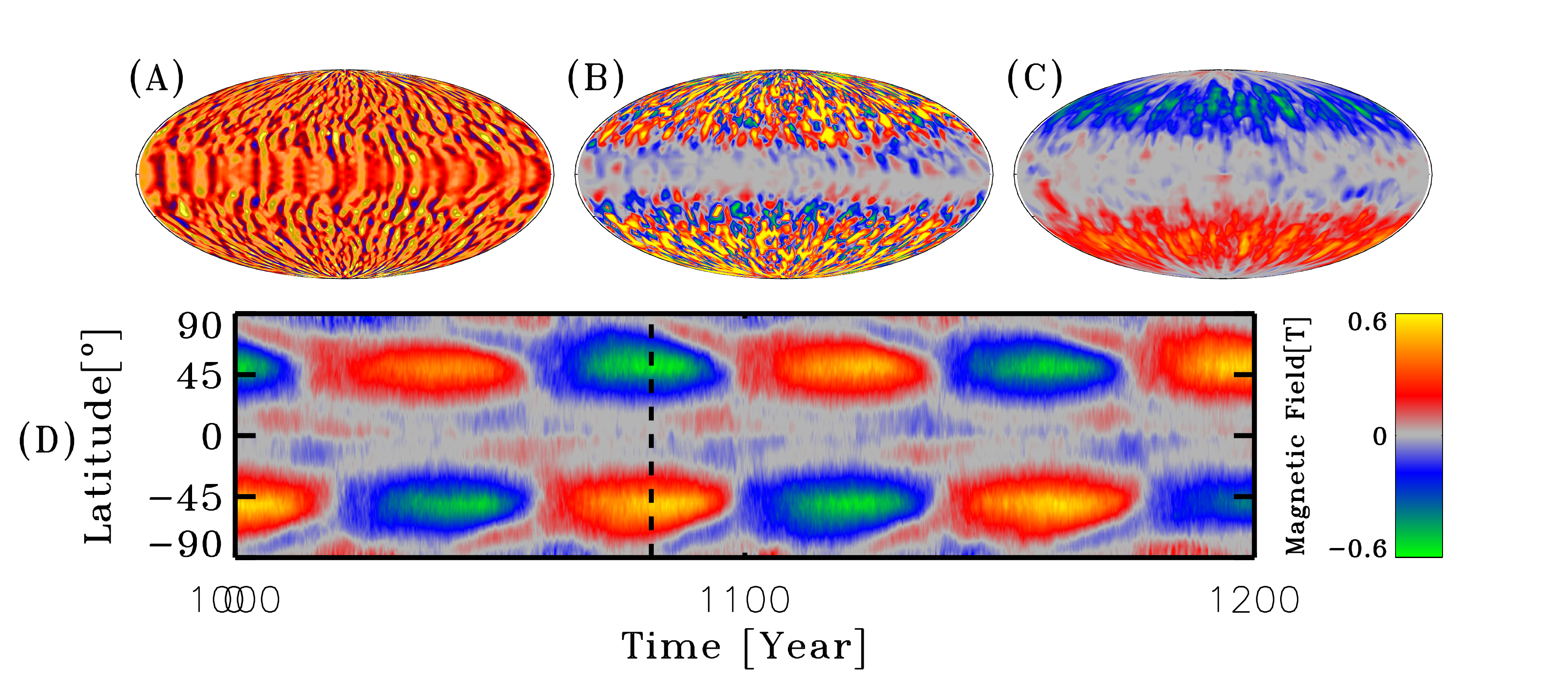}
\caption[Magnetic fields and cycle in EULAG-MHD]
{Snapshots in Mollweide projection of (A) the radial flow component, (B) the radial magnetic component, and (C) toroidal magnetic component, as produced by the the EULAG-MHD ``millenium simulation'' used in the foregoing analysis. Panels A and B are constructed on a spherical shell at mid-depth in the convection zone ($r/R=0.843$) while panel C is constructed at the base of the convecting layers ($r/R=0.711$). Panel D shows a time-latitude ``butterfly'' diagram of the zonally-averaged toroidal component, also constructed at $r/R=0.711$, and illustrates the large-scale magnetic cycle developing in the simulation. The vertical dashed line indicates the epoch from which the snaphots A through C have been extracted, corresponding to the peak of a magnetic cycle.}
\label{fig:structure1}
\end{figure}

Figure \ref{fig:structure1} illustrates some characteristics of the magnetic field building up in our simulation. Panel A and B show, respectively, a Mollweide projection of the radial flow and magnetic field components on a spherical shell at $r/R=0.843$, near the middle of the convectively unstable layers. The magnetic field is quite turbulent, and develops on the same scale as the convective flow, as one would expect considering that convection is the primary inductive flow operating here. 
Figure \ref{fig:structure1}C shows a Mollweide projection of the toroidal (zonally-directed) magnetic component at the base of the convectively unstable layers. The magnetic field is still quite turbulent, but large-scale organization is now also clearly apparent with two strong zonally-oriented
magnetic field bands having built up at mid-latitudes with opposite polarities in each hemisphere. Figure \ref{fig:structure1}D shows a time-latitude ``butterfly'' diagram of the zonally-averaged toroidal magnetic component at the base of the convecting layers, spanning a 200$\,$yr time period. The large-scale magnetic cycle characterizing the axisymmetric magnetic component shows up prominently on such a plot, with its $\sim 40\,$yr half-period, antisymmetry about the equatorial plane, and good hemispheric synchrony.

Here, and in what follows, zonal averages are indicated by angular brackets and computed directly from the simulation output as, e.g., 
\begin{equation}
\langle B_\phi \rangle(r,\theta,t) = \frac{1}{2\pi}\int_0^{2\pi}B_\phi(r,\theta,\phi,t){\rm d}\phi~.
\label{eq:zonalav}
\end{equation}
What we will refer to as non-axisymmetric components, denoted by primes, are then obtained by subtracting such zonal averages from the total flow (${\bf u}$) and magnetic field (${\bf B}$) produced by the simulation:
\begin{equation}
{\bf u}^\prime(r,\theta,\phi,t)=
{\bf u}(r,\theta,\phi,t)- \langle {\bf u} \rangle(r,\theta,t)~,\qquad
{\bf B}^\prime(r,\theta,\phi,t)=
{\bf B}(r,\theta,\phi,t)- \langle {\bf B} \rangle(r,\theta,t)~,\qquad
\label{eq:nonaxi}
\end{equation}

\begin{figure}[hp]
\includegraphics[width=\linewidth]{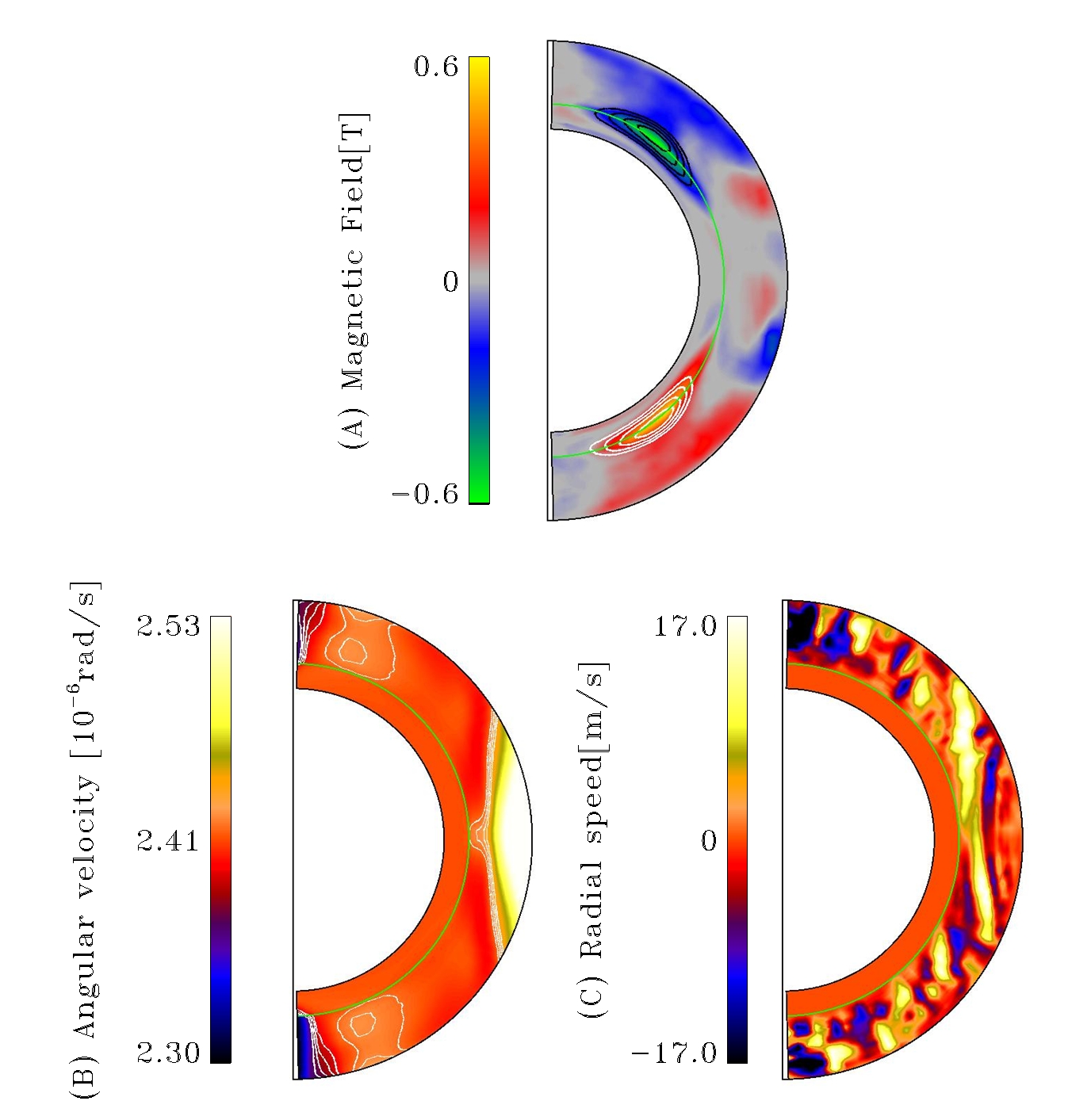}
\caption[Depth structure of flow and magnetic fields in EULAG-MHD]
{Meridional $(r,\theta)$ plane representations of (A) the zonally-averaged toroidal (color scale) and poloidal (black and white contours, field lines oriented clockwise and anticlockwise respectively)
magnetic components at a cycle maximum, (B) the rotational angular velocity averaged zonally and temporally over the full length of the simulation, and (C) a snapshot of the radial flow component at a fixed longitude. The black vertical line is the rotation axis, and the green line indicates the base of the convective zone, at $r/R=0.711$.}
\label{fig:structure}
\end{figure}

Figure \ref{fig:structure} shows yet another view of the same simulation data, this time in the form of meridional slices: zonally-averaged toroidal field snapshot at cycle maximum in A; angular rotational velocity averaged over the whole simulation in panel B; and a meridional slice (fixed longitude) snapshot of the radial flow component at cycle maximum in C. The axisymmetric large-scale magnetic field in (A) is seen to peak immediately beneath the base of the convection zone (green circular arc), and remains largely contained inward of a cylinder aligned with the rotation axis and tangent to the equatorial base of the convection zone. The large-scale field is organized there in the form of two strongly magnetized (peaking at $\simeq 0.5\,$T) bands showing a strong degree of twist, the poloidal and toroidal components being of similar magnitude, as quantified by the ratio
\begin{equation}
\frac{\max \left (\langle{B_\phi}\rangle  \right )_\phi}{\max \left (\langle{B_\theta}\rangle  \right )_\phi + \max \left (\langle{B_r}\rangle  \right )_\phi}~,
\end{equation}
which reaches values in the range $[0.6,1]$ at cycle maxima. A well-organized toroidal component is also present throughout the bulk of the convecting layers, although with weaker amplitude, reaching $\sim 0.1\,T$ in mid-latitudes at cycle maximum.

The differential rotation shown on Figure \ref{fig:structure}B is solar-like, in that it is characterized by significant equatorial acceleration and polar deceleration, both vanishing in the stable layer across a thin shear layer straddling the base of the convective envelope. In this simulation, the pole-to-equator contrast in angular velocity is actually too small by a factor of $\sim 3$ as compared to the sun, and the angular velocity iso-contours at low latitudes show too strong an alignment with the rotation axis. This reflects the strong influence of rotation on convective cells and rolls, which is also apparent in the radial flow snapshot on panel C, where the upflows and downflows tend to be radially-oriented at mid- to high-latitudes, but show strong deviations from the radial direction at low latitudes. In fact, outside of the aforementioned tangent cylinder, convection is organized in the form of a longitudinal stack of convective rolls elongated parallel to the rotation axis. 
This is a robust and common feature of solar convection simulations, whether purely hydrodynamical or magnetohydrodynamical, when operating in this parameter regime \citep[see][]{2009AnRFM..41..317M}. 
These structures may be related to the so-called giant convective cells which have been postulated to exist within the solar convection, and for which indirect observational evidence continues to accumulate \citep[see][and references therein]{Hathaway:2013fb,2014ApJ...784L..32M}. 

Other noteworthy features of this simulation include a well-defined dipole moment, solar-like rotational torsional oscillations and cyclic modulation of convective energy transport 
\citep[see][for further discussion of these features]{Beaudoin:2013eq,2013ApJ...777L..29C,Passos:2014kx}. 
However, as an analog of the sun and its magnetic cycle, the simulation produces a large-scale magnetic field that peaks at too high latitudes compared to the Sun and fails to exhibit equatorward propagation. The cycle period is also nearly four times larger than what is observed. Nonetheless, the nonlinear interactions of flow and magnetic fields are captured in a dynamically consistent manner at all spatial and temporal scaled resolved in the simulation. We therefore proceed with (cautious) confidence.

\subsection{The tachocline and overshoot layer}

In the EULAG-MHD millenium simulation analyzed herein, significant downward pumping and accumulation of magnetic fields takes place at the base of the convecting layers. This is a robust characteristic of numerical simulations of turbulent convection in density-stratified environments, and can be traced to the topological asymmetry between strong narrow downflows of cold fluid, and the gentler broader upflows of warm fluid \citep[see, e.g.,][]{Tobias:2001ho} 
As can be seen on Figure \ref{fig:structure}, the magnetic field accumulates and peaks
in the outer reaches of the underlying stably stratified fluid layers. This behavior is generally consistent with prevalent views of sunspot formation, which posit the storage of toroidal magnetic flux ropes in the overshoot layer, prior to their buoyant destabilization and rise to the photosphere (\citet[][]{Fan:2009ww}; but see \citet[][]{Stenflo:2012aa,Nelson:2013fa} for alternative viewpoints). 

The background stratification used in our EULAG-MHD simulations is characterized by a very strongly stable stratification in the fluid layers underlying the convection zone. More specifically, we use a layered polytropic model, with index varying from a value 1.5 at the base of the convecting layers ($r/R=0.711$), up to 2.5 at the base of the domain ($r/R=0.604$). In conjunction with the very low diffusivities provided by EULAG-MHD's advection scheme in absence of strong velocity and magnetic shear, this naturally leads to the buildup of a thin overshoot layer. This is illustrated on Figure \ref{fig:zonesA}, showing the depth variation of the kinetic energy $E_{Kr}$ associated with the radial component of the flow (solid line) and the root mean square latitudinal deviation $\Delta\Omega$ of the zonally-averaged plasma angular velocity (dash-dotted line), both integrated over spherical shells and time-averaged over the whole simulation:

\begin{figure}[h]
\includegraphics[width=\linewidth]{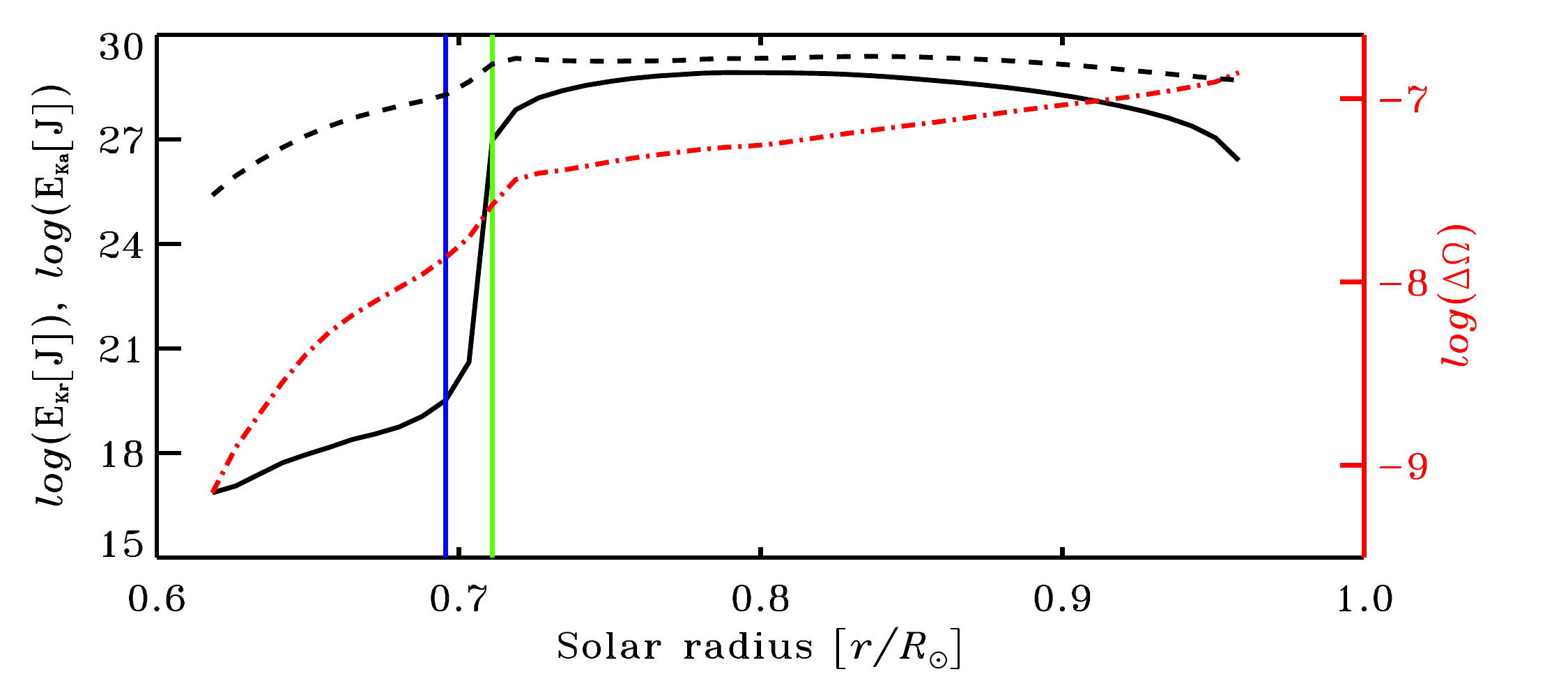}
\caption[Kinetic energy at every radii]
{The solid and dashed lines show the depth variation of the kinetic energy associated respectively with radial and horizontal (latitudinal and zonal) fluid motions, integrated on concentric spherical shells of thickness $\Delta r/R=0.023$. Note that the logarithmic scale on the left covers 15 orders of magnitude. The dash-dotted line shows the RMS latitudinal angular velocity deviation, as defined through eq.~(\ref{eq:DO1})--(\ref{eq:DO2}), also on a logarithmic scale (right axis). The green vertical line indicates the transition from superadiabadic to subadiabatic stratification, i.e., the base of the convection zone. The vertical blue line indicates the base of the overshoot layer, as defined on the basis of the kinetic energy profile (see text).}
\label{fig:zonesA}
\end{figure}

\begin{equation}
E_{K_r} \left( r \right) = \frac{\Delta r}{2T}\int_0^T\int_{0}^{2\pi } \int_0^\pi
(u_r^\prime)^2(r,\theta,\phi,t)\rho(r)r^2 \sin \theta {\rm d}\theta {\rm d}\phi {\rm d}t ~,
\label{eq:EKr}
\end{equation}
\begin{equation}
\Delta \Omega \left( r \right) = \sqrt{\left\langle {\left( {\Omega \left( {r,\theta } \right) - \bar \Omega \left( r \right)} \right)^2 } \right\rangle _\theta} ~,
\label{eq:DO1}
\end{equation}
where
\begin{equation}
\Omega(r,\theta)={\frac{1}{2\pi r\sin\theta}}{\int_0^{2\pi}}
u_\phi(r,\theta,\phi,t) {\rm d}\phi~,
\label{eq:DO2}
\end{equation}
and $\bar\Omega(r)$ denotes spatial averaging over a spherical shell of radius $r$, and $\Delta r$ is a shell thickness corresponding to the radial grid interval. Note that because significant rotational torsional oscillations develop in this simulation \citep[see][]{Beaudoin:2013eq}, 
the calculation of $\Delta\Omega$ and $\bar\Omega$ are based on a full time-average over the whole simulation.

The kinetic energy of radial fluid motions remains roughly constant through the bulk of the convecting layers, but falls precipitously below the base of the convecting layers (vertical green line on Figure \ref{fig:zonesA}), dropping by eight orders of magnitude by $r/R=0.69$ (vertical blue line), followed by slower decrease by another two orders of magnitude down to the base of the stable layer. The finite thickness of the layer across which the kinetic energy drops to zero is due to convective overshoot; the upward-directed buoyancy force below the convectively unstable layers does not decelerate instantaneously the strong, convective 
downflows entering the stable layer from above, so that convective mixing
persists in a thin layer underlying the convectively unstable fluid layers.
We opted to define our ``overshoot layer'' as the depth interval over which the first sharp drop is taking place, i.e., $0.696\leq r/R\leq 0.711$. Henceforth what we refer to as ``stable layer'' thus covers the depth range $0.604\leq r/R\leq 0.696$. The significant, two orders-of-magnitude decrease in $E_{Kr}$ as the top of the simulation domain is approached from below reflects our (conventional) choice of upper boundary condition on the flow for such simulations, namely stress-free and impenetrable.

The latitudinal differential rotation, as measured by $\Delta\Omega$ (dash-dotted line on Figure \ref{fig:zonesA}), is roughly constant in the bottom half of the convecting layers, and also drops in the stable layers, albeit more slowly than the flow kinetic energy. This is due primarily to large-scale magnetic torques contributing to radial fluxes of angular momentum down to $r/R\simeq 0.65$ in this simulation \citep[see Figure 6 in][]{Beaudoin:2013eq}. 
Despite a drop by a factor of $\sim 3$ across the overshoot layer, a weak but significant latitudinal differential rotation remains present in the outer half of our stable layer, with which is also associated a small but significant radial shear peaking at polar and equatorial latitudes. Note that the decrease in latitudinal differential rotation begins already within the convectively unstable layers, consistent with helioseismic inversions indicating that the solar tachocline straddles the base of the convection zone \citep[e.g.,][]{Howe:2009aa}. 
Its thickness, as inferred from Figure \ref{fig:zonesA}, is also consistent with helioseismology results, indicating a solar tachocline thickness no larger than $r/R=0.04$ \citep{Charbonneau:1999es}. 
The slow increase of $\Delta\Omega$ in the outer half of the convecting layers can be traced to the strong equatorial differential rotation building up there (see Figure \ref{fig:structure}B).

The dashed line on Figure \ref{fig:zonesA} shows the depth variation of the kinetic energy $E_{Ka}$ associated with the non-axisymmetric horizontal (i.e., latitudinal and zonal) component of the flow, plotted on the same logarithmic scale as $E_{Kr}$, again integrated over spherical shells and time-averaged over the whole simulation, similarly to our definition of $E_{Kr}$ in eq.~(\ref{eq:EKr}):
\begin{equation}
E_{K_a} \left( r \right) = \frac{\Delta r}{2T}\int_0^T\int_0^{2\pi}\int_0^\pi
(u_\theta^\prime)^2 + (u_\phi^\prime)^2) \rho(r)r^2 \sin\theta
{\rm d}\theta {\rm d}\phi {\rm d}t ~.
\end{equation}

The kinetic energy of the non-axisymmetric horizontal components also stays roughly constant through the bulk of the convecting layers, as does $E_{Kr}$. It also undergoes a sharp drop in the overshoot layer, but only by two orders of magnitude, as opposed to the 15 order-or-magnitude drop observed in $E_{Kr}$. A further, gradual decrease by another three orders-or-magnitude takes place between the base of the overshoot layer and the bottom of the domain. This indicates the presence of predominantly ``horizontal'' fluid motions, i.e. fluid motions constrained to constant radius spherical shells, sustained in the stable layers. This represents a first hint of dynamical effects developing in the stable layer, distinct from the simple buoyant deceleration of convective downflows. We will revisit this issue in \S\ref{sec:instabEvidence}, but as a needed preamble we first examine in greater detail the physical mechanism(s) leading to magnetic field accumulation and amplification in the stable layers.

\section{Origin of the magnetic fields in the stable layer}
\label{sec:magStable}

\begin{figure}[h]
\includegraphics[width=\linewidth]{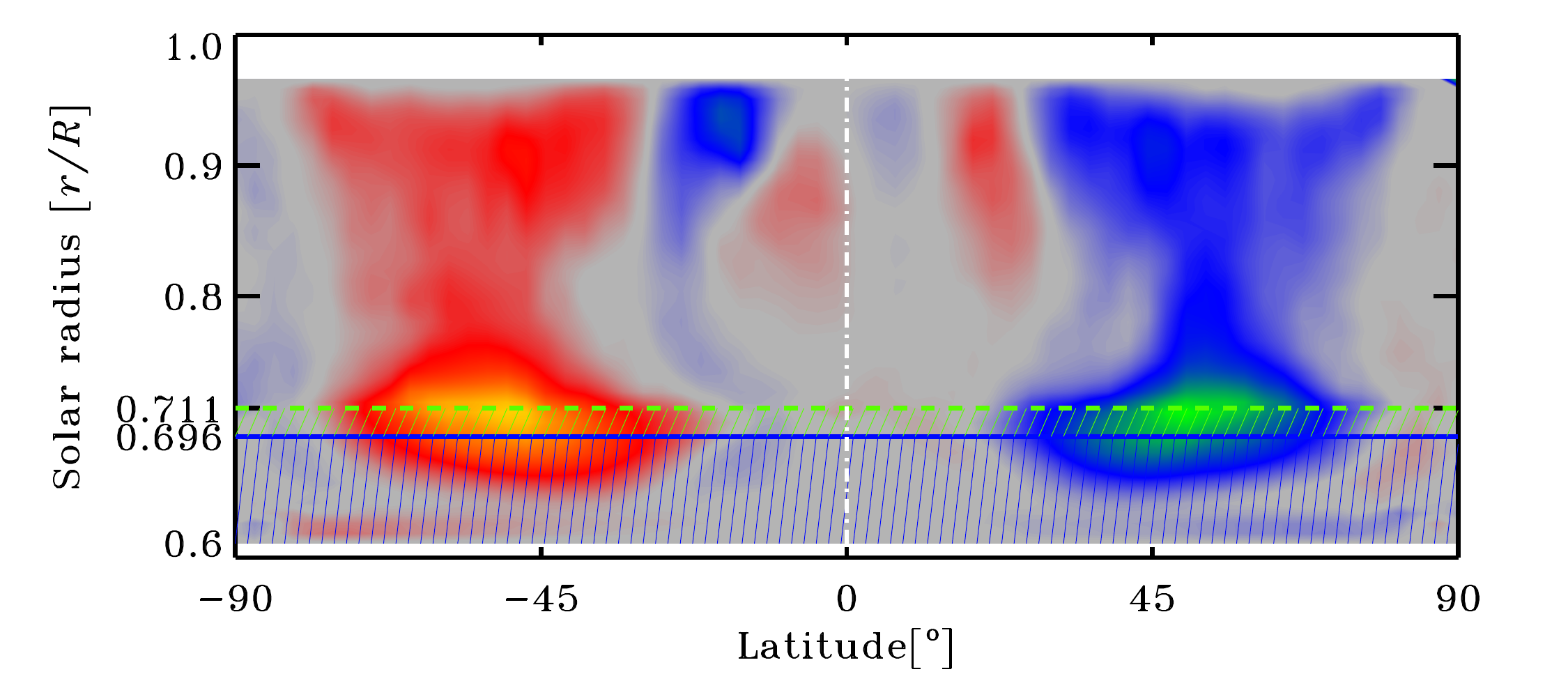}
\caption[Toroidal magnetic field and EULAG-MHD zone separation]
{Zonally-averaged toroidal magnetic component at typical magnetic cycle peak, as on Figure \ref{fig:structure}A, but now plotted in a cartesian radius-latitude plane. The green and blue horizontal lines delineate the limits of the overshoot layer (cf. Figure \ref{fig:zonesA}), with the blue hatched area defining the stable zone.}
\label{fig:zonesB}
\end{figure}

Figure \ref{fig:zonesB} offers a more detailed look at the localisation of the large-scale magnetic field bands in relation to the overshoot and tachocline layers defined on the basis of Figure \ref{fig:zonesA}. This is the same zonally-averaged toroidal magnetic field at cycle peak displayed on Figure \ref{fig:structure}A, but plotted this time in a cartesian radius-latitude plane. The toroidal field bands peak at $\simeq 0.5\,$T in the overshoot layer, but also extend halfway down into the stable zone with significant amplitude ($\geq 0.1\,$T).

It is interesting to note that at high radii near the equator, a different mean toroidal magnetic field develops at non-negligible levels. This magnetic field appears to be generated by a second, local dynamo tapping into the strong radial shear in the equatorial
outer half of the convection zone, and undergoing polarity reversals on a much
shorter period ($\sim 2\,$yr) than the primary deep-seated dynamo cycle \citep[for more in this dual cycle behavior, see][]{Beaudoin:2015aa}. 

One may rightfully wonder whether the strong magnetic fields building up in the overshoot layer and underlying stable layer results only from accumulation of magnetic fields pumped downward from the convecting layers, or if local inductive effects are also contributing. As a first step towards answering this question, we first use the simulation output to compute, at each grid point, the Poynting flux:
\begin{equation}
{\bf S} = \frac{1}{{\mu _0 }} {\bf E} \times {\bf B}=
\frac{1}{{\mu _0 }}( {\bf u} \times {\bf B} ) \times {\bf B}~,
\end{equation}
and integrate its radial component on spherical shells:
\begin{equation}
P(r,t)  = \int_0^{2\pi} \int_0^\pi 
S_r( r,\theta ,\phi ,t)r^2 \sin\theta {\rm d}\theta {\rm d}\phi~.
\label{eq:PF}
\end{equation}

\begin{figure}[hp]
\includegraphics[width=\linewidth]{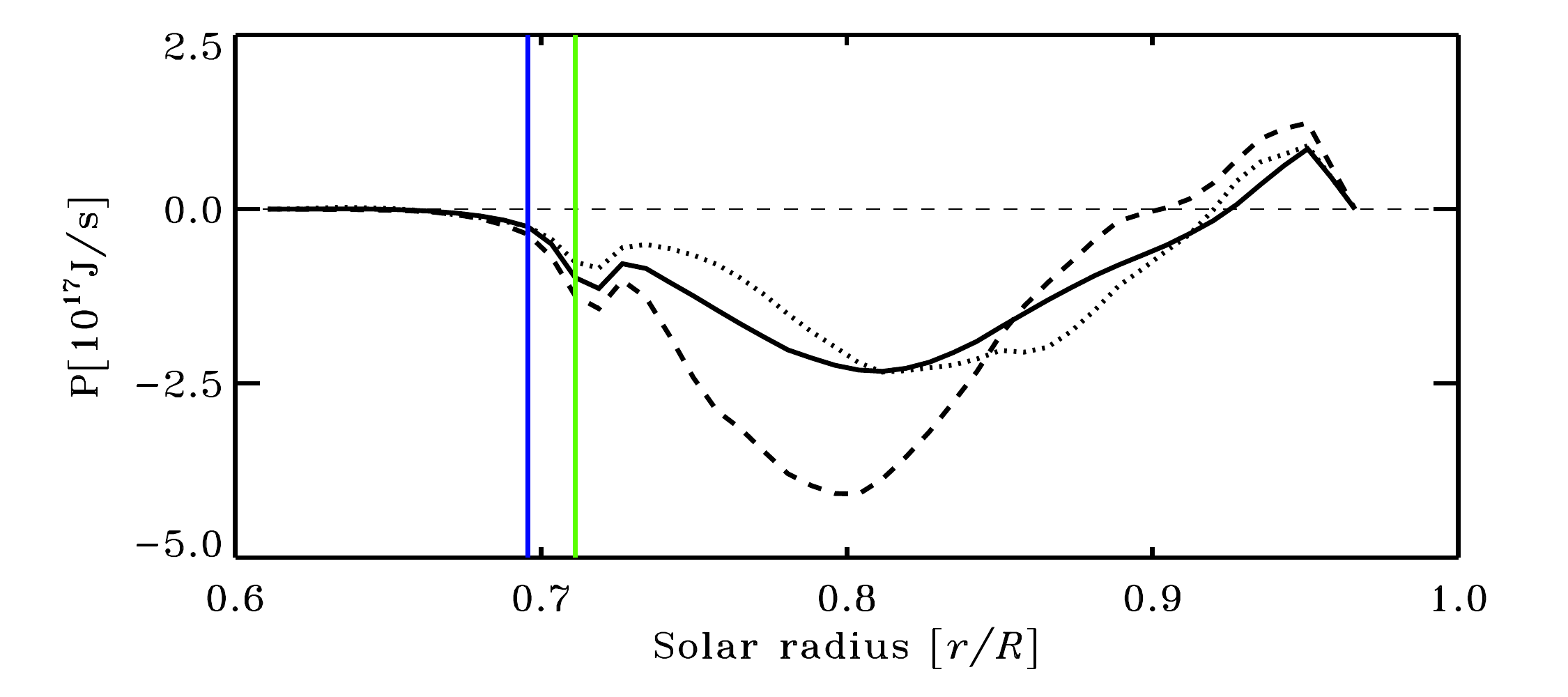}
\caption[Pointing flux]
{Depth variation of the radial Poynting flux integrated over concentric spherical shells (viz.~eq.~\ref{eq:PF}). The solid line shows an average over the whole duration of the simulation, and the dashed and dotted are computed from snapshots at cycle maximum and minimum respectively. The green and blue vertical line segments are carried over from Figure \ref{fig:zonesA}, and delimit the radial extent of the overshoot layer.}
\label{fig:Poynting}
\end{figure}

Figure \ref{fig:Poynting} shows the depth variation of this quantity, time-averaged over the whole simulation (solid line), as well as equivalent profiles extracted at an epoch of magnetic cycle peak (dashed line) and polarity reversal (dotted line). In all cases the integrated radial Poynting flux is negative at all depths except in subsurface layers, as a consequence of our imposed impenetrable upper boundary condition. Magnetic field accumulation is expected wherever the divergence of the Poynting flux is negative, which here is generally the case in the depth range $r/R\leq 0.8$ at all phases of the cycle, consistent with the field accumulation seen on Figure \ref{fig:zonesB}. Note also how the Poynting flux reaches deep into the stable zone, despite the rapid disappearance of turbulent fluid motion below the overshoot layer.

\begin{figure}[hp]
\includegraphics[width=\linewidth]{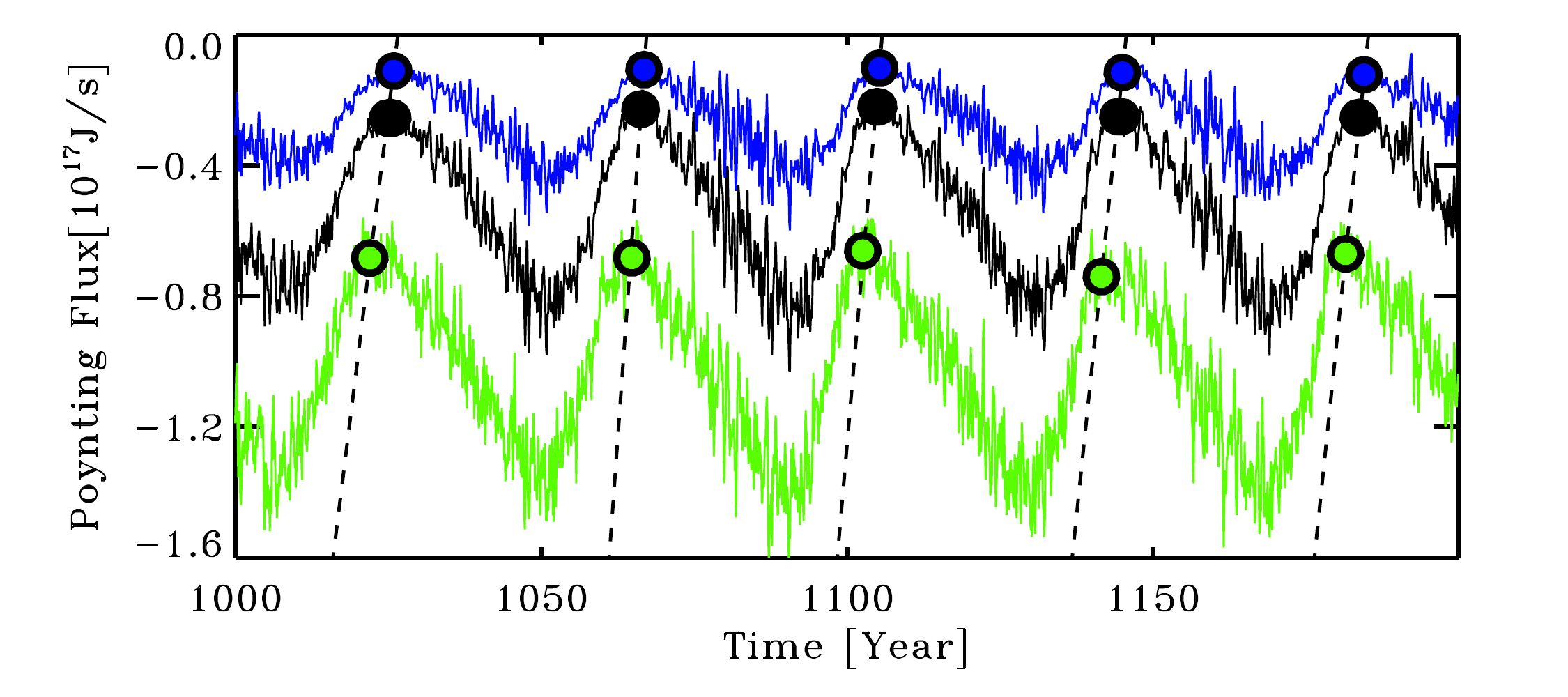}
\caption[Pointing flux in time integrated over 3 different shells]
{Poynting flux integrated over 3 shells (from top to bottom: blue, $r/R=0.696$; black, $r/R=0.704$; green, $r/R=0.711$). The oblique dashed lines connect the minima of each cycle in the time sequence, indicated by correspondingly colored large solid dots.}
\label{fig:TPA}
\end{figure}

The radial Poynting flux also shows a strong dependence on the phase of the large-scale magnetic cycle developing in the simulation, varying cyclically and approximately in phase with the magnetic cycle in the bulk of the convecting layer. In the overshoot and stable layers, however, a temporal lag is observed, as illustrated on Figure \ref{fig:TPA}. This Figure shows time series of the shell-integrated radial Poynting flux at three depths in the overshoot and stable layers. Note how the Poynting flux peaks later and later as one moves progressively deeper below the base of the convection zone. This is consistent with the Poynting flux being driven primarily from above, presumably through induction of the large-scale magnetic field by turbulent dynamo action within the convecting layers.

\begin{figure}[hp]
\includegraphics[width=\linewidth]{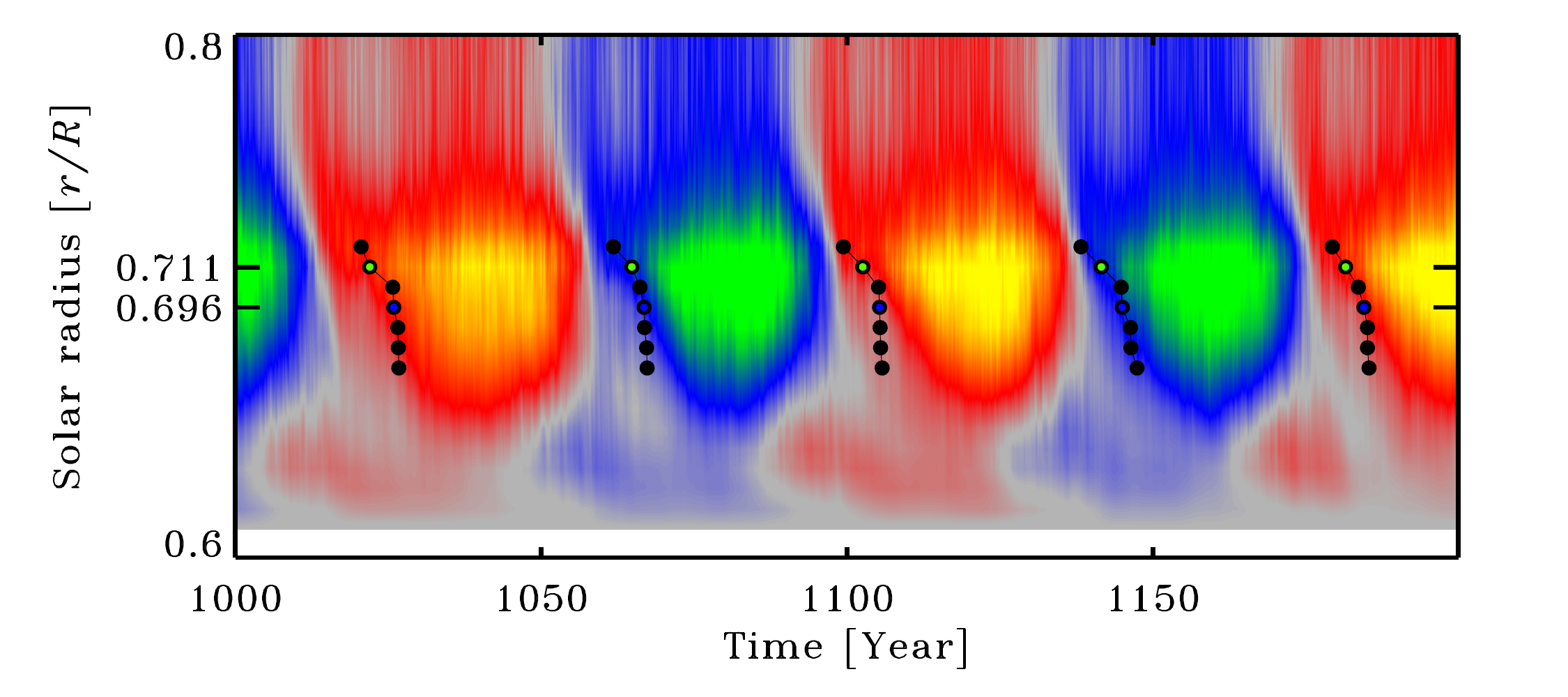}
\caption[Radii-Time representation of the Poynting flux]
{Zonally-averaged toroidal magnetic component plotted as a time-radii slice at 46.1$^\circ$ latitude in the N-hemisphere, over a $\sim$200$\,$yr segment of the millenium simulation. The solid dots indicate epochs of minimal shell-integrated Poynting flux at a few depths going from the base of the convection zone, through the overshoot layer and into the stable zone. The green and blue dots identify the minima in the correspondingly colored time series on Figure \ref{fig:TPA}.}
\label{fig:TPB}
\end{figure}

Figure \ref{fig:TPB} shows the location of the minima in the shell-integrated Poynting flux in a time-radius diagram, superimposed on a time-radius slice of the zonally-averaged toroidal magnetic component at 46.1$^\circ$ latitude. The minima occur shortly after the polarity reversal, consistent with the subsequent buildup of the deep toroidal field at and below the base of the convecting layers. The downwards slant across the overshoot layer is the direct counterpart of the dashed oblique lines on Figure \ref{fig:TPA}. Note, however, the break of slope at the base of the overshoot layer, suggestive of an additional ---and possibly local--- inductive and/or transport process contributing to the spatiotemporal variations of the Poynting flux.

We now compute the total magnetic energy content ($E_M$) in the overshoot and stable layers by direct integration of the simulation output in the depth range $0.61\leq r/R\leq 0.711$, at an epoch of cycle maximum:
\begin{equation}
E_M={1\over 2\mu_0}\int_0^{2\pi}\int_0^\pi\int_{0.61}^{0.711}
{\bf B}^2 r^2\sin\theta {\rm d}r{\rm d}\theta {\rm d}\phi~.
\end{equation}
The timescale for field accumulation beneath the convection zone is then directly
obtained by dividing this quantity by the value of the shell-integrated Poynting flux at $r/R=0.711$:
\begin{equation}
\tau={E_M\over P}~.
\label{eq:timescale}
\end{equation}
The resulting numerical value is $\tau \simeq 8\,$yr, i.e., about 20\% of the $\sim 40\,$yr magnetic half-cycle period in our simulation. This suggests that sufficient magnetic energy is being provided by downward turbulent pumping to destroy the deep toroidal flux bands of the preceding half-cycle and rebuild a band of opposite polarity in each hemisphere. This result is thus consistent with a minimalistic scenario whereby the buildup and reversal of the toroidal flux bands \emph{below} the convection zone are merely a passive side-effect of global dynamo action \emph{within} the convection zone \citep[on this point, see also][]{2013ApJ...778...11M}. 

However, $\tau$ is but a rough estimate, and other mechanism are potentially at play. As noted already in the context of Figure \ref{fig:zonesA}, some level of latitudinal differential rotation, both radial and latitudinal, persists across the overshoot layer and well within the stable zone. With a significant large-scale poloidal magnetic component also present, shearing by differential rotation can contribute to the induction of a toroidal component. This process is captured by the zonal component of the MHD induction equation, which in the ideal limit and for axisymmetric large-scale magnetic field and differential rotation reduces to
\begin{equation}
{\frac{\partial B_\phi}{\partial t}}= r\sin\theta B_r{\frac{\partial\Omega}{\partial r}}+\sin\theta B_\theta{\frac{\partial\Omega}{\partial \theta}}
\label{eq:zonalind}
\end{equation}
Separately integrating $B_\phi$ and each term on the RHS of the above expression over the radial and latitudinal extent of the toroidal field at a time of solar maximum allows to compute characteristic timescales for induction by the radial and latitudinal shear, again by simply dividing the first by the other two. Both of these timescales end up at $\simeq 30\,$yr; while significantly larger than the timescale (\ref{eq:timescale}) associated with the downward Poynting flux, these are still comparable to the half-cycle period for $\simeq 40\,$yr. One can but conclude that differential rotation shear contributes significantly to toroidal field induction in the overshoot and stable layers.

On the basis of this new estimate, an additional factor is thus added to the minimal scenario outlined above: the buildup and reversal of the deep-seated magnetic field bands occurs primarily through turbulent pumping from above, with additional amplification provided by the differential rotation shear. We could stop here, but one key piece of evidence compels us to push our analysis further, namely the existence of significant horizontal kinetic energy in the stable layer (dashed line on Figure \ref{fig:zonesA}).

\section{Searching for the signature of instabilities}
\label{sec:instabEvidence}

\cch{As noted earlier, the kinetic energy of the non-axisymmetric horizontal flow components ($E_{Ka}$ shown in Figure \ref{fig:zonesA}) remains at significantly high values well into the stable layer, before finally dropping abruptly by several orders of magnitudes at the lower boundary, primarily a consequence of the artificial friction terms introduced there.
These horizontal flows persist well below the depth of convective overshoot
(blue vertical line on Figure \ref{fig:zonesA}),
and so cannot be directly driven by the overlying convection. One
possibility is that they arise through the development of one (or more)
local fluid instability.}

\cch{The solar tachocline is the site of significant latitudinal and radial
rotational shear, and its upper reaches include the overshoot layer, which
is only weakly subadiabatic. These characteristics are known to be
conducive to the development of a wide array of fluid instabilities
\citep[see, e.g.,][\S 2.7]{2000stro.book.....T}.
The unmagnetized tachocline is stable according
to the classical Rayleigh criterion \citep{Watson:1981uc,Charbonneau:1999kw},
but it has recently been shown
to be prone to the development of the baroclinic instability
\citep{2014ApJ...787...60G}. 
However, it is unlikely that we are capturing
this instability in our simulation, for a number of reasons.
First, the radial shear in the simulation's ``tachocline'' is significantly
weaker than in the real sun \citep[see Figure 2 in][]{Beaudoin:2013eq}.
Second, the Newtonian
cooling term used to drive convection damps out baroclinicity
by forcing the temperature stratification towards a spherically-symmetric
state which is strongly subadiabatic in the convectively stable layers.
Third, the magnetic fields
building up in the stable layers of our simulation
reach the strength $\sim 0.2\,$T at which they can stabilize
the baroclinic instability, at least according to the recent
local stability analyses of \citet{2015ApJ...801...22G}.}

\cch{On the other hand,
there also exist circumstances under which the presence
of magnetic fields can also have a destabilizing effect. This is the
case with the magnetoshear instability, to which we now turn.
}

\subsection{Magnetoshear instability}

Numerous analytic, semi-analytic and numerical calculations have identified two-dimensional MHD instabilities that can become excited in stably stratified regions of the solar interior in the presence of latitudinal differential rotation and large-scale magnetic field. Particularly pertinent to our simulation is the so-called joint MHD instability first investigated by \citet{Gilman:1997cr} \citep[see also][]{1999ApJ...512..417D,2001ApJ...551..536D,2003ApJ...596..680D,Gilman:2007fy}. 
This instability develops in stably stratified environments in the presence of axisymmetric latitudinal differential rotation and large-scale toroidal magnetic fields, i.e., the situation prevailing in the outer reaches of the stable layer in our simulation. The instability planforms are 2D, i.e., they develop on spherical surfaces, and the most unstable modes have low azimuthal wavenumbers, $m=1$ or $2$, as magnetic tension provides a restoring force that strongly suppresses higher wavenumber modes \citep[see][and references therein]{Gilman:2007fy}. 
Depending on the cross-hemispheric phasing of the instability planform, the global development of the instability can lead to magnetic reconnection across the equator 
\citep[``clamshell instability''; see, e.g.,][]{2003ApJ...582.1190C},
or the two toroidal flux systems can both tilt while remaining parallel to one another across the hemispheres (``tipping instability'').

The magnetoshear instabilities tap into both the kinetic energy of the differential rotation, as well as the magnetic energy of the large-scale toroidal magnetic field. It is a ``joint'' instability, in that both ingredients are required for the instability to grow. The growth rate $s$ of the most unstable mode is typically some fraction of the Alfv\'en time based on the toroidal magnetic field strength:
\begin{equation}
s\sim \frac{L}{u_A},\qquad u_A=\sqrt{\langle B_\phi\rangle^2\over \mu_0\rho}~,
\end{equation}
where $L$ is a typical length scale for the toroidal magnetic field. Using $L=R$, $\langle B_\phi\rangle=0.5\,T$, and $\rho=10^{-2}\,$kg m$^{-3}$, appropriate for the upper part of the stable layer in our simulation, yields $s\simeq 2\,$yr; with a half-cycle period of some 40$\,$yr, the instability would presumably have sufficient time to fully develop in the course of a magnetic cycle.

The nonlinear development of this instability has been investigated by \citet{Miesch:2007fp} 
using a non-linear 2D shallow-water MHD model of a stably stratified thin shell domain identified with the tachocline. 
He showed that, provided an external (to his thin shell domain) source is available to replenish the differential rotation and poloidal magnetic field destroyed by the nonlinear saturation of the instability, the latter can be sustained with energy being cyclically exchanged between the axisymmetric and non-axisymmetric magnetic components. Although our EULAG-MHD simulation is 3D, the strongly subadiabatic stratification in the stable layer restricts plasma flows to spherical shells, so that Miesch's model could be applicable to each concentric spherical shell. We therefore follow \citet{Miesch:2007fp} 
by defining instability proxies through the energies associated with the non-axisymmetric magnetic field (Non-Axisymmetric Magnetic Energy, hereafter NAME), and the magnetic energy associated with the axisymmetric toroidal component (Toroidal Field Magnetic Energy, hereafter TFME). For our 3D simulation, these are defined through the following volume integrals:
\begin{equation}
{\rm NAME}(t) =
\frac{1}{2\mu_0}\int_{0.611}^{0.696} \int_{[0,\pi/2]}^{[\pi/2,\pi]} \int_0^{2\pi}
({\bf B}^\prime)^2r^2 \sin\theta {\rm d}\phi {\rm d}\theta {\rm d}r~,
\label{eq:NAME}
\end{equation}
\begin{equation}
{\rm TFME}(t) =
\frac{1}{2\mu_0}\int_{0.611}^{0.696} \int_{[0,\pi/2]}^{[\pi/2,\pi]} \int_0^{2\pi}
\langle B_\phi\rangle^2 r^2 \sin\theta{\rm d}\phi {\rm d}\theta {\rm d}r~.
\label{eq:TFME}
\end{equation}

Unlike in \citet{Miesch:2007fp}
's model, here the poloidal field is generated autonomously; it will therefore prove useful to define an analog of (\ref{eq:TFME}) for the axisymmetric poloidal field:
\begin{equation}
{\rm PFME}(t) =
\frac{1}{2\mu_0}\int_{0.611}^{0.696 } \int_{[0,\pi/2]}^{[\pi/2,\pi]} \int_0^{2\pi}
(\langle B_r\rangle^2 +\langle B_\theta\rangle^2) r^2 \sin\theta{\rm d}\phi {\rm d}\theta {\rm d}r~.
\label{eq:PFME}
\end{equation}

This proxy turns out to be largely dominated by the contribution from the latitudinal component $\langle B_\theta\rangle$. Note that in order to ensure cleaner proxy time series, we only integrate over the stable the layer, excluding the overshoot layers. Moreover, as detailed in \citet{Passos:2014kx}, 
magnetic cycles in the EULAG-MHD millenium simulation can show small but significant phase lag between hemispheres; all of the above proxy integrals are therefore computed separately for the Northern $(0\leq\theta\leq\pi/2)$ and Southern ($\pi/2\leq\theta\leq \pi$) hemispheres.

\begin{figure}[htbp]
\includegraphics[width=\linewidth]{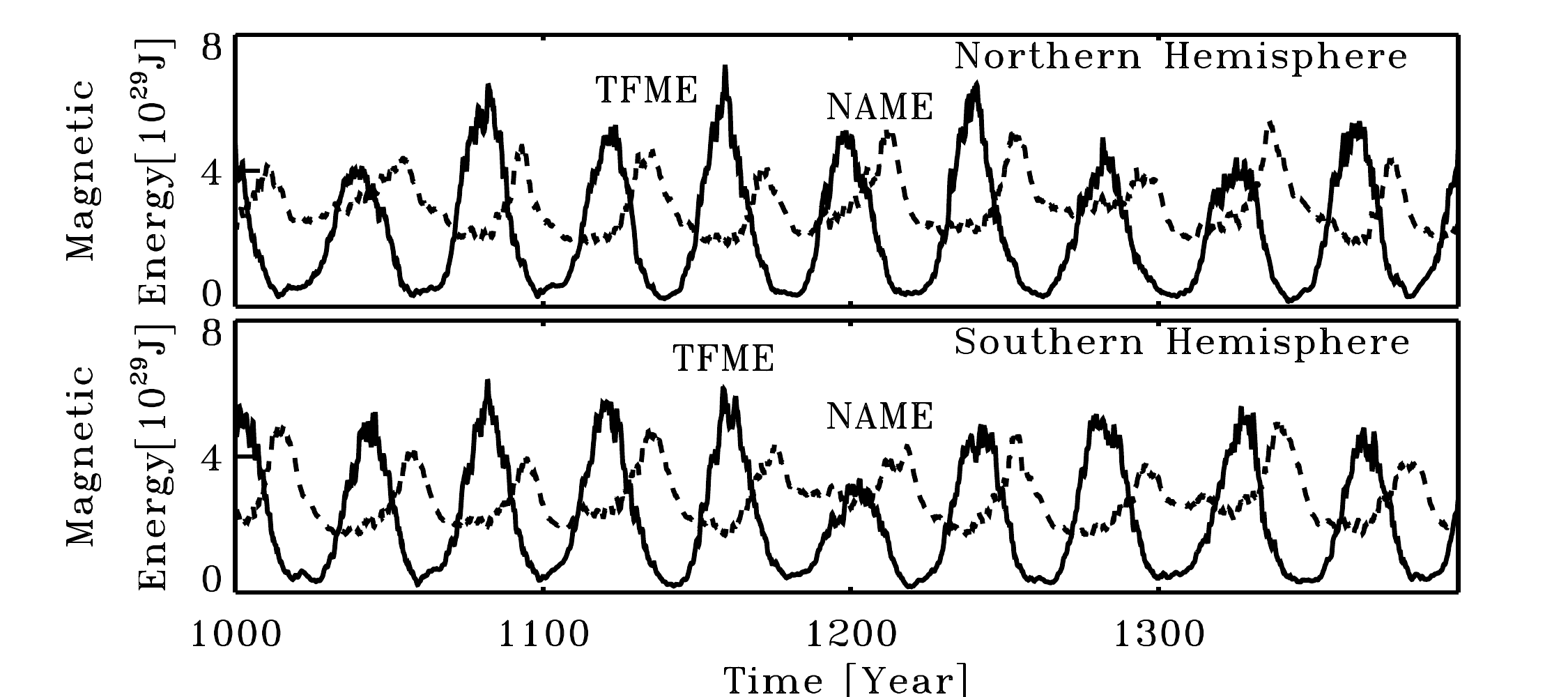}
\caption[Instability proxy and magnetic cycle proxy]
{Time series of the volume-integrated magnetic energy contained in the axisymmetric toroidal field (TFME; solid line) and the total non-axisymmetric field (NAME; dotted line) in the stable zone of the EULAG-MHD millenium simulation, for the Northern (top) and Southern (bottom) hemispheres. Compare to Figure 2 in \citet{Miesch:2007fp}. 
}
\label{fig:proxyTime}
\end{figure}

Figure \ref{fig:proxyTime} shows a representative 400yr segment of the hemispheric time series for the TFME (solid line) and NAME (dashed line) proxies. Both proxies wax and wane cyclically while maintaining a well-defined phase lag, with NAME peaking in the late descending phase of TFME, and the onset of the growth phase of NAME almost always occurring near the peak of TFME. This is remarkably similar to the pattern characterizing the forced 2D simulations of \citet{Miesch:2007fp} 
(compare his Figure 2 to Figure \ref{fig:proxyTime} herein). This suggests ---of course without strictly proving--- that we are observing in the millenium simulation the same type of magnetoshear instability investigated by \citet{Miesch:2007fp}; 
however, no external forcing is imposed here, as the latitudinal differential is being maintained by Reynolds stresses within the convection zone, and the poloidal field is naturally replenished by turbulent dynamo action within the convecting layers and subsequent downward pumping of the magnetic field. In particular, the roughly similar amplitudes of variation in TFME and NAME, $\sim 4\times 10^{29}J$, are consistent with the cyclic exchange of energy between the axisymmetric and non-axisymmetric magnetic components characterizing the forced shallow water simulations of \citet{Miesch:2007fp}. 

Under flux freezing, any global displacement of the toroidal field bands due to the magnetoshear instability ---whether developing in its clamshell or tipping variants--- must be accompanied by a corresponding latitudinally-oriented flow constrained to spherical shells. Figure \ref{fig:zonesA} already indicates the predominance of horizontal flows in the stable layer, and the characterization of such flow can provide further evidence that the instability is indeed operating. Accordingly, we now define two measures of flow kinetic energy similar to those used for magnetic energy: the total kinetic energy of the poloidal (PKE) and toroidal (TKE) components. As before, we integrate separately the Northern and Southern hemisphere:
\begin{equation}
{\rm PKE}(t) =
\frac{1}{2}\int_{0.611}^{0.696} \int_{[0,\pi/2]}^{[\pi/2,\pi]} \int_0^{2\pi}
\left( u_\theta^2 + u_r^2 \right) \rho r^2 \sin\theta {\rm d}\phi {\rm d}\theta {\rm d}r~.
\label{eq:PKE}
\end{equation}
\begin{equation}
{\rm TKE}(t) =
\frac{1}{2}\int_{0.611}^{0.696} \int_{[0,\pi/2]}^{[\pi/2,\pi]} \int_0^{2\pi}
u_\phi^2 \rho r^2 \sin\theta {\rm d}\phi {\rm d}\theta {\rm d}r~.
\label{eq:TKE}
\end{equation}

Note that we use here the total flow components, but since very little axisymmetric meridional flow develops in the stable layer, PKE is essentially the flow equivalent of NAME. Moreover, as with the PFME proxy, PKE is dominated by the latitudinal component $u_\theta$.

\begin{figure}[hp]
\includegraphics[width=\linewidth]{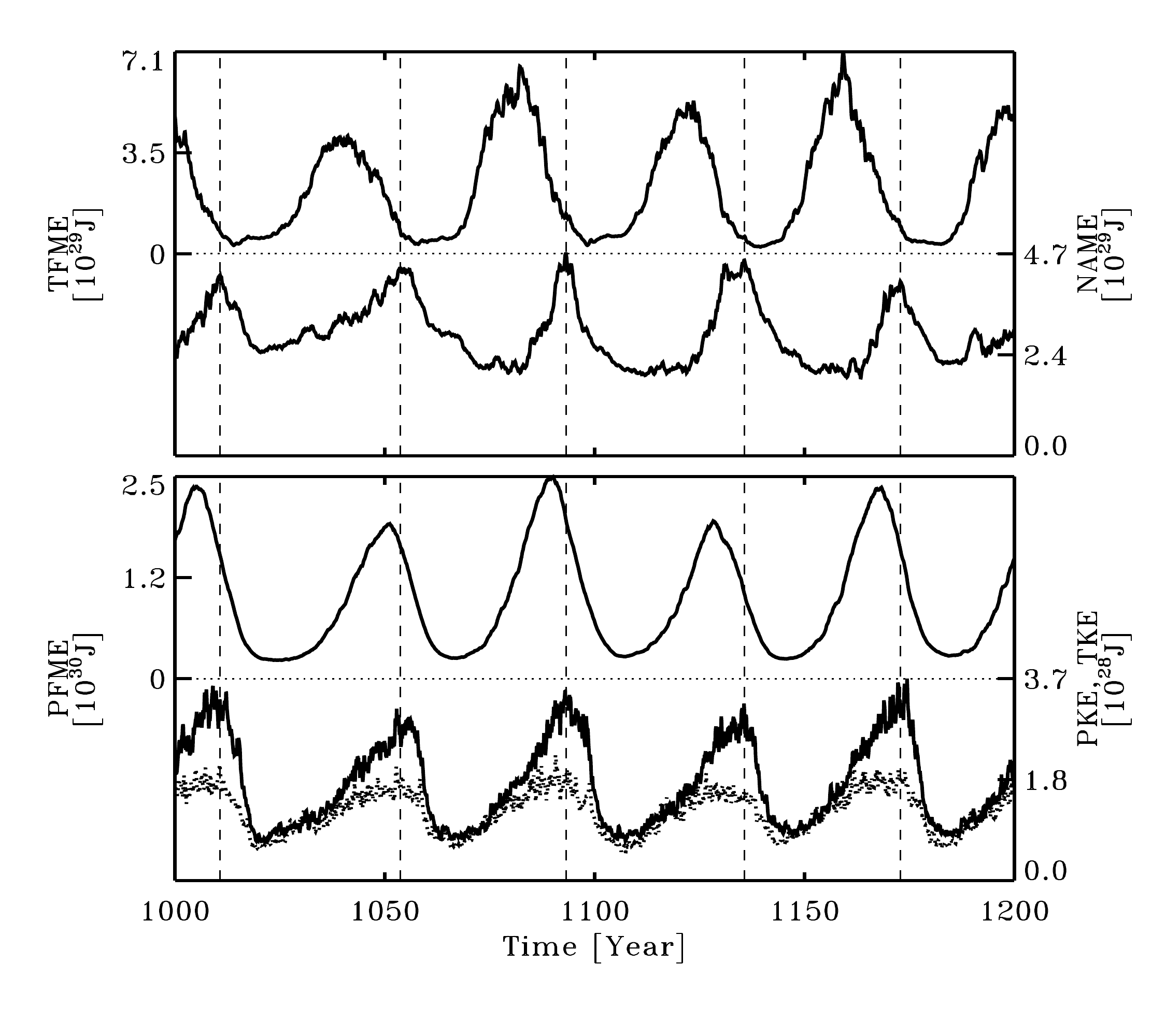}
\caption[PFME and rotational energy]
{The top panel shows a 200yr$\,$ closeup of Figure \ref{fig:proxyTime} for the N-hemisphere, and the bottom shows the corresponding time series for the PKE (solid line) and TKE (dotted line) flow energy proxies defined through eqs.~(\ref{eq:PKE})--(\ref{eq:TKE}). The poloidal magnetic proxy PFME (viz.~eq.~\ref{eq:PFME}) is also shown. The vertical dashed lines indicate the epochs of peak NAME. Note the in-phase variations of PKE and TKE with NAME.}
\label{fig:proxyECin}
\end{figure}

Figure \ref{fig:proxyECin} shows these two time series over a restricted 200$\,$yr temporal span, together with the magnetic proxies defined earlier, all in the Northern hemisphere. The dashed vertical lines indicate the epoch of peak NAME in the descending phase of each of the five magnetic cycles developing over the time period covered. Note that our (axisymmetric) poloidal field magnetic energy PFME also exhibits cyclic behavior, in response to dynamo action in the overlying convecting fluid layers, unlike in the simulation of \citet{Miesch:2007fp} 
where the poloidal (latitudinal) field is imposed externally and remains constant.

Examination of Figure \ref{fig:proxyECin} reveals that both PKE and TKE evolve in phase with NAME, with peak-to-through variations $\sim 0.3$ and $0.1\times 10^{29}\,$J, over an order of magnitude lower than the corresponding variations in TFME. Indeed, the combined rise of NAME, TKE, and PKE adds up to $\sim 2.5 \times 10^{29} J$, of the same order but still comfortably smaller than the associated $\sim 4 \times 10^{19} J$ drop in TFME. The in-phase variation of PKE with NAME suggests that the horizontal flow develops simultaneously with the growth of
the non-axisymmetric magnetic component, which is the pattern expected for the clamshell or tipping variants of the magnetoshear instability. As plasma is displaced away from or towards the rotation axis, conservation of angular momentum causes zonal deceleration or acceleration, leading to a growth of TKE in phase with PKE, as is indeed observed on Figure \ref{fig:proxyECin}.

\subsection{Spectral decomposition}
\label{sec:spectr-decomp}

In order to properly identify the clamshell --or tipping variants-- of the magnetoshear instability, we decompose the magnetic field on the spherical harmonics in the stable layer. We use the classical vectorial spherical harmonics basis \citep[see][]{Rieutord:1987go,Mathis:2005kz,Strugarek:2013kt} 
\begin{equation}
  \mathbf{B} = \sum_{l,m} \alpha_{l,m}\mathbf{R}^{m}_{l} + \beta_{l,m}\mathbf{S}^{m}_{l} + \gamma_{l}^{m}\mathbf{T}^{m}_{l} \, ,
  \label{eq:B_decomp}
\end{equation}
where the orthogonal basis $\left(\mathbf{R}^{m}_{l},\mathbf{S}^{m}_{l},\mathbf{T}^{m}_{l}\right)$ is defined by
\begin{equation*}
  \left\{
  \begin{array}{lcl}
    \mathbf{R}^{m}_{l} &=& Y^{m}_{l} \mathbf{e}_{r} \\
    \mathbf{S}^{m}_{l} &=& \boldsymbol{\nabla}_{\perp} Y^{m}_{l} = \partial_{\theta} Y^{m}_{l}\mathbf{e}_{\theta} + \frac{1}{\sin{\theta}}\partial_{\varphi} Y^{m}_{l}\mathbf{e}_{\varphi}\\
    \mathbf{T}^{m}_{l} &=& \boldsymbol{\nabla}_{\perp}\times \mathbf{R}^{m}_{l} = \frac{1}{\sin{\theta}}\partial_{\varphi} Y^{m}_{l}\mathbf{e}_{\theta} -\partial_{\theta} Y^{m}_{l}\mathbf{e}_{\varphi}
  \end{array}
  \right.  \, ,
\end{equation*}
with $Y^{m}_{l}$ being the classical spherical harmonics. At a given depth, the magnetic energy spectrum can be decomposed along the spherical harmonic degrees $m$ by
\begin{equation}
  \label{eq:ME_decomp}
  {\rm ME}_{m}(r,t) = \frac{V_{r}}{2\mu_{0}}\sum_{l\ge m} \left|\alpha_{l,m}\right|^{2} + l(l+1)\left(\left|\beta_{l,m}\right|^{2} + \left|\gamma_{l,m}\right|^{2}\right)\, ,
\end{equation}
where $V_{r}$ is the local volume of the spherical shell centered on the cell located at depth $r$, included for dimensional consistency.

\begin{figure}[hp]
\includegraphics[width=\linewidth]{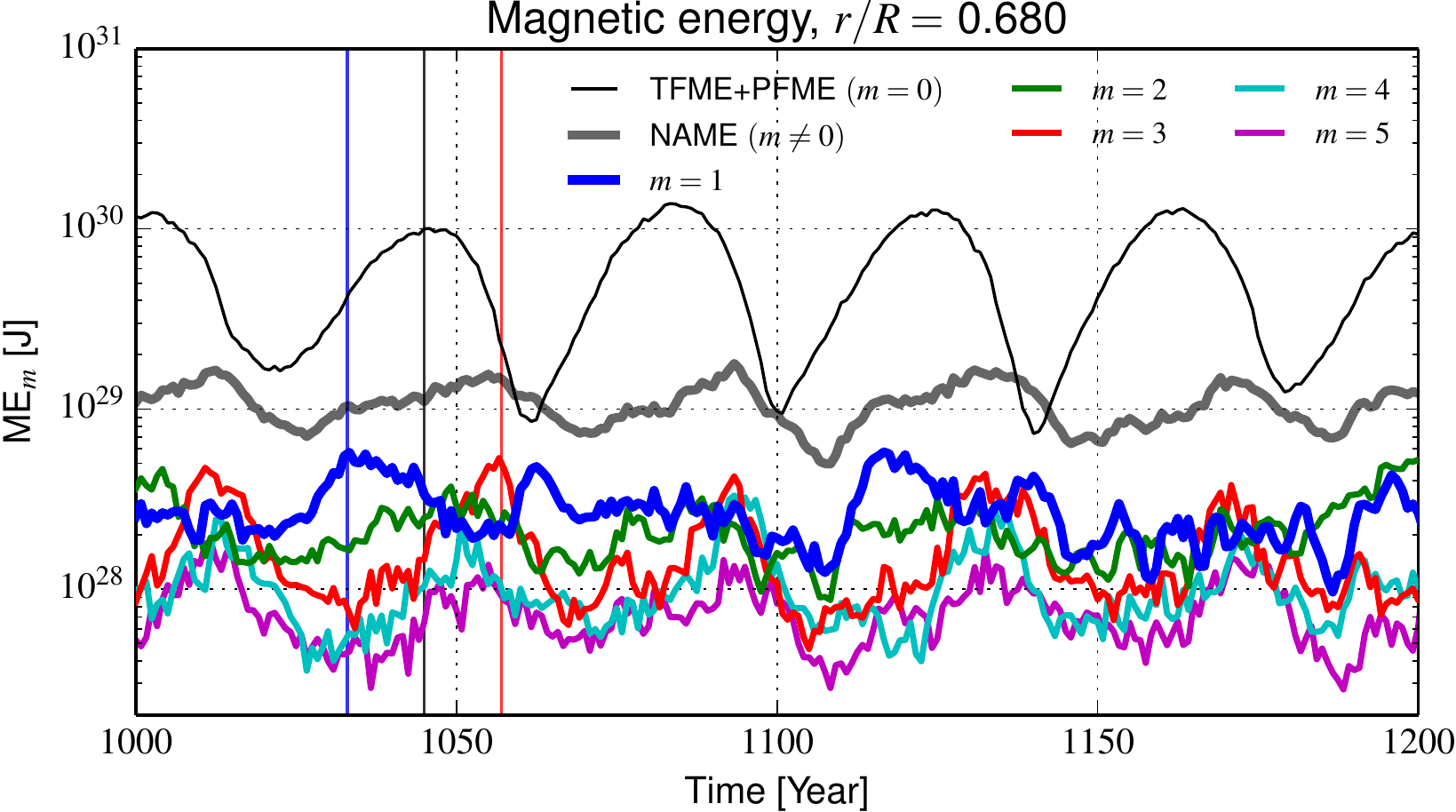}
\caption[Spectrum evolution along m]{Magnetic energy evolution at $r/R=0.680$, decomposed over the first spherical harmonics degrees $m$, summed over $l$. The total, axisymmetric ($m=0$) magnetic energy, which is dominated by the PFME contribution, is indicated in black. The non-axisymmetric (NAME, $m\ne 0$) in shown in bold grey. The three vertical lines indicate the time at which the mollweide projection in Figure \ref{fig:moll_clam} were taken.}
\label{fig:sp_m_evol}
\end{figure}

We display the evolution of the magnetic energy for $m\in [0,5]$ in Figure \ref{fig:sp_m_evol} at depth $r/R=0.680$. The magnetic energy for $m=1$ traces the onset of a magnetoshear instability which is synchronized with TFME (see Figure \ref{fig:proxyECin}). The $m=1$ growth is accompanied by an $m=2$ component of the magnetic energy, which saturates later in the cycle, in phase with the higher $m$ components and with NAME. The non-axisymmetric magnetic energy (NAME) cycle maxima are dominated by the $m=3$ or $m=4$ modes, depending on the considered cycle. Once the tipping instability has taken off, magnetic energy stored in the most unstable $m=1$ and $m=2$ modes is expected to be non-linearly transfered to those higher $m$'s due to the simultaneous development of horizontal flows at the same scales. 

\begin{figure}[hp]
\includegraphics[width=\linewidth]{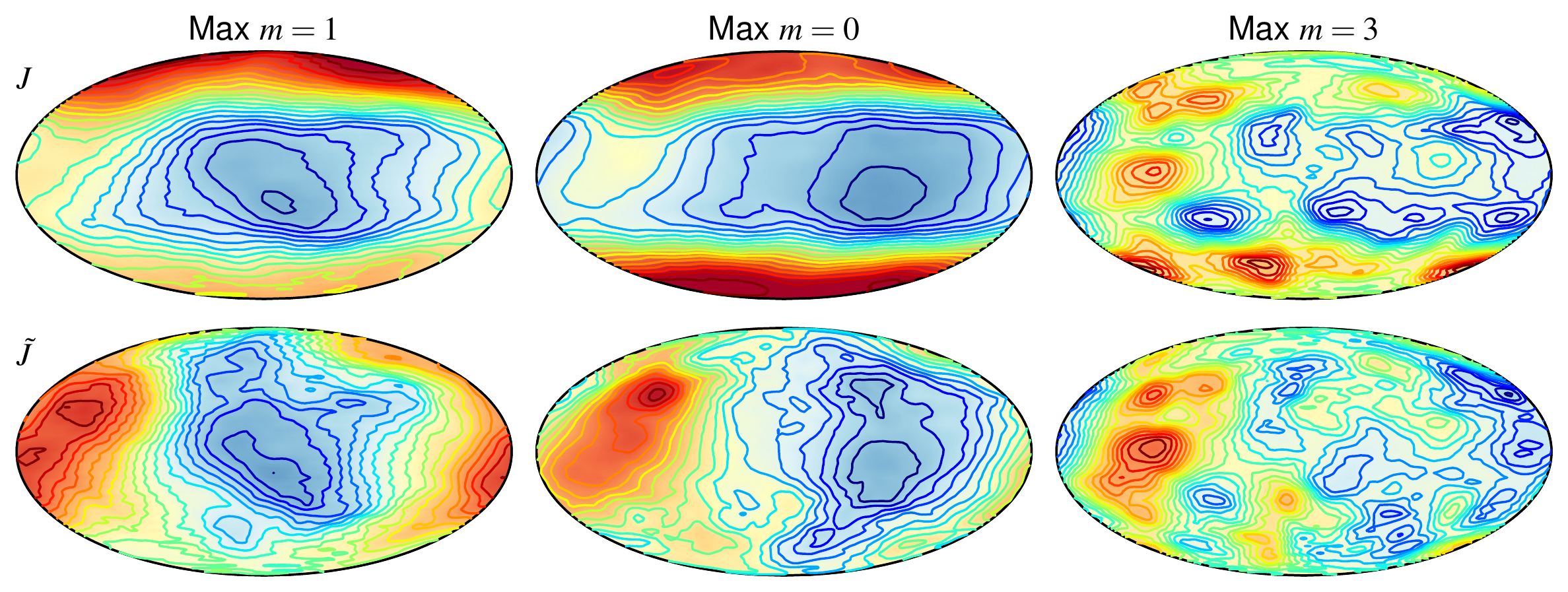}
\caption[Mollweide clam]{Magnetic field lines on the spherical surface $r/R=0.680$, well within the stable layer. From left to right, the diagram are taken at the times labeled by the three vertical lines in Figure \ref{fig:sp_m_evol} corresponding respectively to epochs of $m=1$ maximum, $m=0$ maximum and $m=3$ maximum. The upper diagrams correspond the streamfunction of the full horizontal magnetic field, and the lower diagrams to the streamfunction of the non-axisymmetric horizontal magnetic field. On these Mollweide projections a purely toroidal magnetic field would have all its field lines oriented horizontally. Compare to Figure 3 in \citet{Miesch:2007fp}. 
}
\label{fig:moll_clam}
\end{figure}

In order to geometrically identify the tipping instability that develops on the shells harboring the strong toroidal magnetic fields, we again follow \citet{Miesch:2007fp} 
and visualize magnetic field lines as iso-contours of the magnetic streamfunctions $J$ and $\tilde{J}$, defined such that ${\bf B} \sim \hat {\bf z} \times \nabla J$ and ${\bf B}-\langle{\bf B}\rangle_{\varphi} \sim \hat {\bf z} \times \nabla \tilde{J}$. These fields lines are shown on Figure \ref{fig:moll_clam}, constructed at a depth $r/R=0.680$ well below the overshoot layer at epochs of $m=1$ maximum (left), TFME maximum (middle) and $m=3$ maximum (right).

The epoch of maxima of $m=1$ (left panels) generally occurs concurrently with the peak of TFME, before the maxima of PFME (middle panels). The field lines are thus predominantly axisymmetric and composed of longitudinal field at mid and high latitudes, and latitudinal field near the equator. A hint of global $m=1$ tilting can be seen in the two upper left panels, which are confirmed by the predominance of an $m=1$ structure in the two lower left panels. At the time $m=3$ is maximized it dominates the magnetic energy spectrum and no clear longitudinally aligned field lines can be observed on the right panels. Albeit $m=3$ dominates, the field lines exhibit a complex pattern of mixed $m$ components that populate the magnetic energy spectrum at the peak of NAME.

\begin{figure}[hp]
\includegraphics[width=0.5\linewidth]{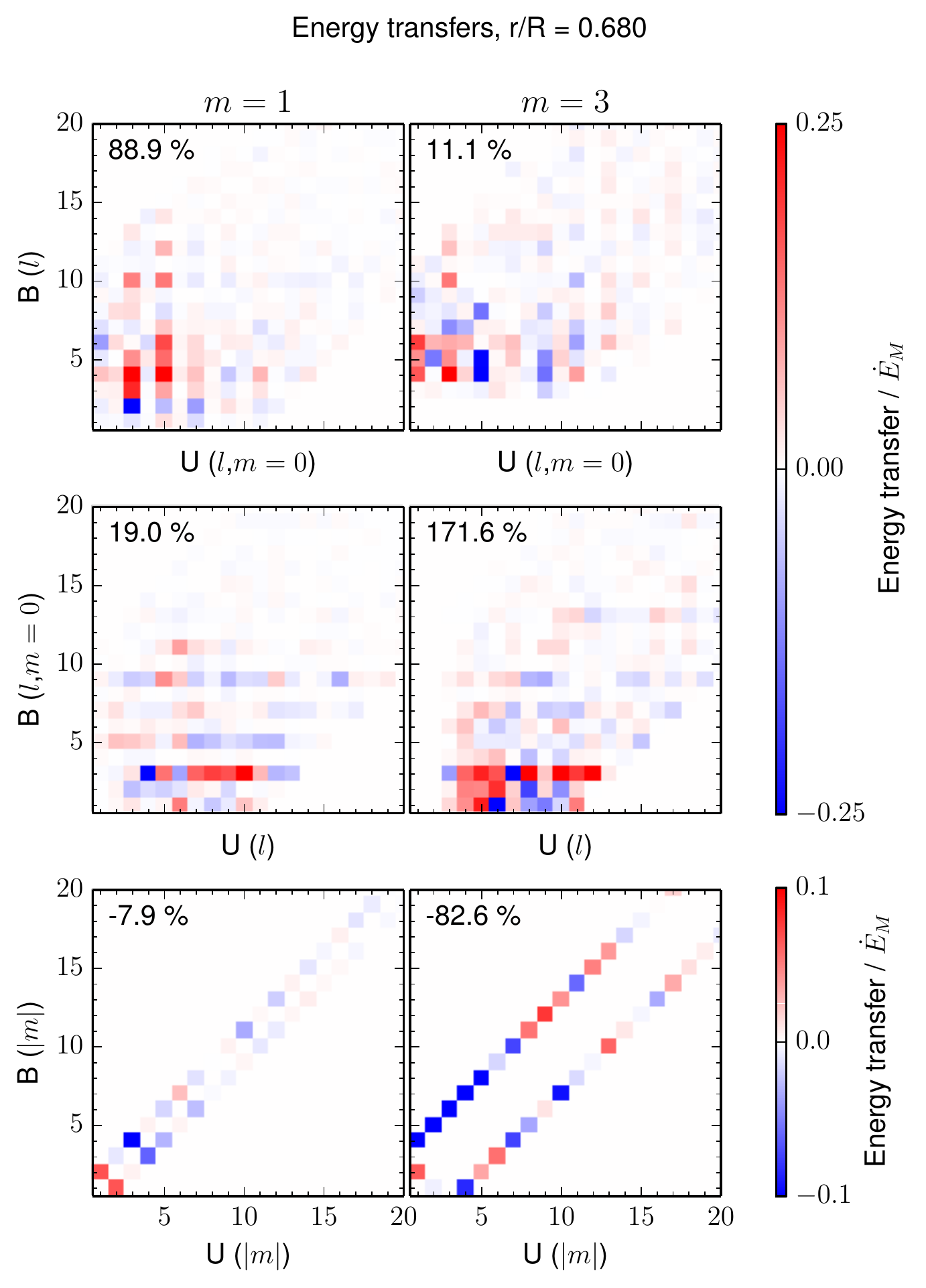}
\caption[Spectral transfer]{\ch{Energy transfer maps towards the $m=1$ (left column) and $m=3$ (right column) components of the magnetic energy spectrum. The transfers are averaged during a typical growth phase of those components, \cch{which are summed over $l$}. The colormap denotes positive (red) and negative (blue) energy transfers and is normalized to the variation of the magnetic energy $\dot{E}_M$. The transfers are separated in three panels in each column, allowing to identify transfers from the axisymmetric ($m=0$) and non-axisymmetric ($m\neq 0$) velocity (abscissas) and magnetic (ordinates) fields. \cch{The upper panels hence show transfers with the axisymmetric differential rotation, and the middle panels transfers with the axisymmetric magnetic field, as a function of $l$. In the lower panels we focus on the fully non-axisymmetric transfers for which we chose the energy conversions as a function of $m$, summed over $l$.} The percentage listed at the upper left in each panel corresponds to the sum over all the contributions in the panel.}}
\label{fig:sp_transfer}
\end{figure}

\ch{Further insight is gained by quantifying the spectral energy transfers leading to the successive growth of the $m=1$ and $m=3$ modes. \cch{During its maxima periods, the $m=1$ mode completely dominates the non-axisymmetric spectrum. We choose to focus on the $m=3$ mode because it provides a typical dominant mode during the NAME maxima periods, and is always one of the few most energetic modes in each cycle (the $m=4$ and $m=5$ modes sometimes dominate the NAME spectrum, but they significantly vary from one cycle to the other).} We follow the procedure detailed in \citet{Strugarek:2013kt} 
to quantify the amount of energy transfered to a given $m$ due to the triadic interactions involving two other spherical harmonics $(l_1,m_1)$ and $(l_2,m_2)$. We display in Figure \ref{fig:sp_transfer} energy transfers maps \cch{(summed over $l$)} towards $m=1$ (left column) and $m=3$ (right column) during their growth phase for a typical cycle. \cch{Hence, the left column was averaged during a different time-window than the right column, to account for the time-lag between the growth periods of the two modes (see also Figure \ref{fig:sp_m_evol}).} Positive (red) and negative (blue) transfers respectively represent a source and a sink of magnetic energy for the chosen $m$. In each panel, the horizontal axis corresponds to \cch{the velocity field spectrum}, 
and the vertical axis to 
\cch{the magnetic field spectrum}. \cch{The contributions are separated in non-axisymmetric ($m\neq 0$) and axisymmetric ($m=0$) fields}, leading to three possible ways to transfer energy. 

\cch{We show in the upper panels the transfers involving the differential rotation, and in the middle panels the transfers involving the large scale magnetic field. The axes correspond to the various $l$ couplings, which allows us to identify the dominant transfers from the differential rotation ($l=3$, $5$, corresponding to the vertical red stripes in the upper panels) and the large-scale magnetic field ($l=3$, corresponding to the horizontal red stripes in the middle panels). We see that a large range of non-axisymmetric scales $l$, coupled to the axisymmetric fields, are involved in the transfer. In the lower panels, we focus on the fully non-axisymmetric couplings and the axes now represent the $m$ modes of the magnetic and velocity fields, summed over $l$. The triadic selection rule naturally leads to diagonal-only transfers in the lower panels. The transfers to higher $m$ mode play little to no role in the case of $m=1$, whereas the $m=3$ mode transfers a significant amount of energy to the $m=4$ to $10$ modes during its growth, as one would expect from a cascade-like process.} The color-scale is saturated at $25\%$ of the total change of the magnetic energy $\dot{E}_M$ during the considered growth period in the upper and middle panels, and to $10\%$ in the lower panels. It appears clearly that the $m=1$ and $m=3$ modes both gain energy from the differential rotation and the large-scale magnetic field, while the non-linear interactions with higher order modes tend to stabilize them. \cch{The total magnetic energy transfer from each channel is quantified by the percentage in the upper left corner of each panel}. The $m=1$ mode hence dominantly grows by receiving energy directly from the large-scale differential rotation, while the $m=3$ mode preferentially draws energy from the axisymmetric large-scale magnetic field at $l=3$. Note that for each column the total is not exactly $100\%$ due to the fact that the transfer maps were averaged during the magnetic energy growth phase. These results are robust for the different growth phases shown in Figure \ref{fig:sp_m_evol}.}

\ch{Using the fully detailed analysis from \citet{Strugarek:2013kt} 
(not shown here), it is further possible to disentangle the origin of the energy transfers with respect to, \textit{e.g.}, the radial or latitudinal structure of the differential rotation. Our simulation shows that the $m=1$ mode dominantly draws energy from the latitudinal differential rotation, while the $m=3$ mode grows by extracting energy from the radial structure of the axisymmetric large-scale magnetic field. As a result, these energy transfers are a compelling evidence of the occurence of an MHD instability akin to the magnetoshear instability in our simulation.
}

The magnetic energy nevertheless largely dominates the energy balance in the stable layers (see Figure \ref{fig:proxyECin}). \ch{The $m=1$ mode grows by receiving energy from both the differential rotation and the large-scale axisymmetric magnetic field}, as a result, it is possible that some other type of MHD instability is instead at play, and we now turn to this possibility.

\section{Digging further: the Tayler instability}
\label{sec:tayler}

The magnetoshear instability investigated by \citet{Miesch:2007fp} 
is far from the only one that can potentially develop in stably stratified, weakly magnetized differentially rotating astrophysical environments. For example, the turbulent stresses generated by the magneto-rotational instability \citep[MRI; see][]{Balbus:1991fi} 
are now believed to dominate the outward transport of angular momentum in weakly magnetized accretion disks. This is a very powerful, local MHD instability, requiring a significant poloidal magnetic component to operate, but in the stellar
interior context it also requires a significant outwardly decreasing
radial differential rotation.
The high-latitude regions of the tachocline satisfy this criterion, and
could thus be the seat of a localized version of this instability
\citep[see, e.g.,][]{2007ApJ...667L.207P,2011MNRAS.411L..26M}.
However, if present throughout the overshoot layer and tachocline, it would lead to significant radial mixing. This appears ruled out here, on the basis of Figure \ref{fig:zonesA} which indicates that fluid motions in the stable layer are strongly restricted to spherical surfaces. \cch{The magnetic buoyancy instability \citep{Parker:1955bb} is also a potential candidate,
but again it would lead to radial mixing, which we do not observe in the
stably stratified layers of our
simulation. Moreover, we likely do not have the spatial resolution required
to capture the formation of thin magnetic flux tube-like structures
conducive to the development of this instability, at least judging
from the much higher resolutions simulations of \citet{Nelson:2013fa}.
To the best of our knowledge, at this writing these remain the only global MHD simulations of
solar convection in which the spontaneous onset of this instability
has been observed.} 

Another class of MHD instabilities, the ``Tayler instabilities'', have also been investigated in detail in the context of various types of large-scale magnetic fields embedded in stellar radiative interiors. Particularly relevant in the present context is the instability of a purely toroidal magnetic field in the ideal MHD limit \citet{Tayler:1973tn,Spruit:1999vt}. 
In the absence of rotation, any such magnetic field  $B_\phi \left(r,\theta \right)$ is unstable, no matter how weak the field is. The instability is most prone to develop close to the magnetic symmetry axis, where $B_\phi = 0$, as no restoring force can resist the magnetic force pointing towards the axis (see, e.g., the discussion in \cite{Spruit:1999vt};  
also \cite{2002A&A...381..923S,Brun:2006bb}). 
The instability can develop both axisymmetric and non-axisymmetric planforms, with the lowest zonal mode (azimuthal wavenumber $m=1$ in a spherical harmonic expansion) usually most unstable in the latter case (Zahn et al.~2007), because these develop the least magnetic tension tending to oppose the instability. As with the magnetoshear instability investigated in the preceding section, the growth rate for the Tayler instability is of the order of the Alfv\'en time. Stability criteria for axisymmetric purely toroidal magnetic fields have been obtained by \citet{1981Ap&SS..75..521G} 
for spherical geometry, and take the form:
\begin{equation}
\frac{1}{4 \pi r^2 \sin^2 \theta} \left[ 2 H_\phi^2 \cos^2 \theta - \sin \theta \cos \theta \frac{\partial H_\phi^2}{\partial \theta} \right] > 0~,\qquad m=0~,
\label{eq:old_m0}
\end{equation}
\begin{equation}
\frac{1}{4 \pi r^2 \sin^2 \theta} \left[ H_\phi^2 \left( m^2 - 2 \cos^2 \theta \right) - \sin \theta \cos \theta \frac{\partial H_\phi^2}{\partial \theta} \right] > 0~,\qquad m=1~,
\label{eq:old_mm}
\end{equation}
where $H_\phi^2 = b_l^2\left(r \right) P_l^2 \left(\cos \theta \right)$. For the purposes of the foregoing analysis we shall simply assume $H_\phi\equiv \langle B_\phi\rangle \left(r,\theta \right)$, which leads to:
\begin{equation}
2 \cos^2\theta - 2 \cos\theta\sin\theta
\frac{\partial \ln\,\langle B_\phi\rangle}{\partial \theta} > 0~,\qquad m=0~,
\label{eq:new_m0}
\end{equation}
\begin{equation}
1 - 2 \cos^2\theta - 2 \cos\theta\sin\theta
\frac{\partial \ln\,\langle B_\phi\rangle}{\partial \theta} > 0~,\qquad m=1~.
\label{eq:new_m1}
\end{equation}

These stability criteria are admittedly obtained in idealized conditions differing significantly from those encountered within our numerical simulation: ideal MHD, no rotation, purely toroidal magnetic field. However, \citet{1985MNRAS.216..139P} 
showed that rotation weakens but does not suppress the instability. Likewise, the instability is also weakened, but not suppressed, in the presence of a large-scale poloidal magnetic field component. In fact, the numerical simulations of \citet{2006A&A...449..451B} 
confirm the generally unstable nature of large-scale magnetic fields in stably stratified, radiative stellar interiors, and also indicate that the most stable large-scale magnetic field configurations have poloidal and toroidal components of comparable strengths \citep{Braithwaite:2006da}. 
Interestingly, and as noted earlier in \S \ref{sec:magnetic}, the strong magnetic field bands building up across the base of the convection zone in our simulation also have toroidal and poloidal large-scale magnetic components of comparable magnitudes.

Of course, the above stability criteria are of limited use in the analysis of a numerical simulation having reached a nonlinearly saturated statistically stationary state. If an instability is indeed operating, what one observes in the simulation is the global flow and field structure resulting from the saturation of the instability, rather than the original unstable background structure on which eqs.~(\ref{eq:new_m0})--(\ref{eq:new_m1}) could be legitimately applied.  In such a situation one would expect the global magnetic field profiles to exhibit marginal stability when computing the stability criteria. Nonetheless, \citet{Rogers:2011cx} 
found evidence for the development of the axisymmetric ($m=0$) form of the Tayler instability, in a manner consistent with eq.~(\ref{eq:old_m0}), in her 2D axisymmetric MHD simulations of the solar radiative interior including a poloidal magnetic field and imposed differential rotation \citep[a forcing setup somewhat as in][]{Miesch:2007fp}. 
Her simulations show that even in the absence of significant bona fide dynamo action, as long as the poloidal component persists the differential rotation can  induce a toroidal component which, upon becoming unstable to the axisymmetric Tayler instability, undergoes polarity reversals (see her Figure 6). Even closer to the physical situation of interest here, \citet{Brun:2006bb} 
performed 3D MHD simulation of the solar tachocline in which they also observe the development of what they suggest is the non-axisymmetric form of the Tayler-like instabilities, persisting at all depths and particularly prominent in the vicinity of the polar axis, as expected of this instability.

\begin{figure}[hp]
\includegraphics[width=\linewidth]{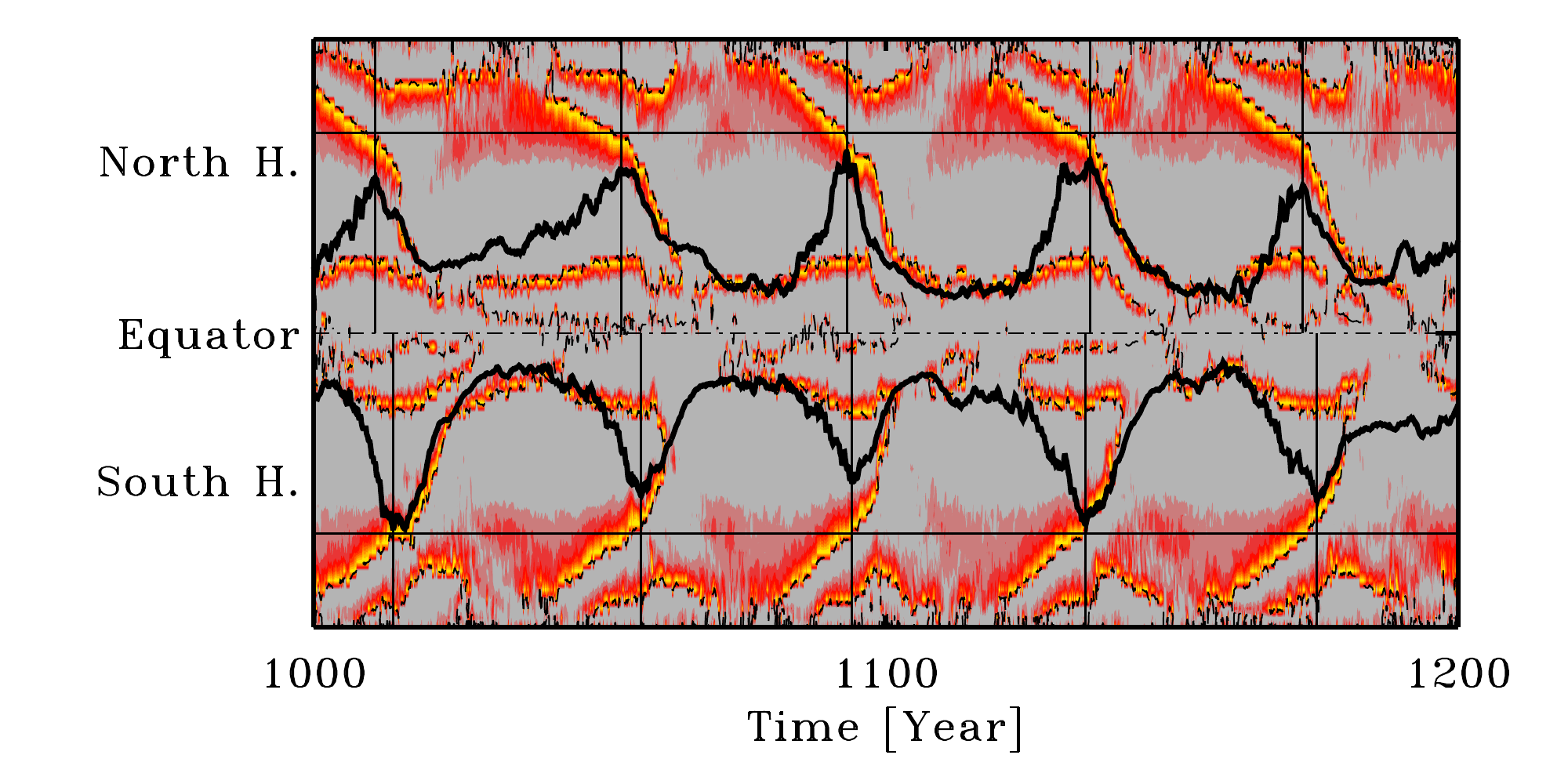}
\caption[Instability criterion compared to the instability proxy]
{Time-latitude color rendering of the $m=1$ instability criterion as given by eq.~(\ref{eq:new_m1}), based on the magnetic field extracted at depth $r/R=0.696$, at the top of the stable layer. Areas in gray are stable, and red-to-yellow represents increased instability. This 200$\,$yr time segment spans five magnetic half-cycles, and is the same as for the time-latitude of the axisymmetric toroidal field plotted on Figure \ref{fig:structure1}D. The thick black lines are the NAME time series, that for the S-hemisphere plotted as negative, and the thin vertical black line indicate epochs of peak NAME. The thin horizontal black lines are drawn at $\pm 61^\circ$ latitude, corresponding to the poleward extent of the strong ($\langle B_\phi\rangle\geq 0.1\,$T) magnetic field bands building up in the stable layer. The dashed black line shows the $\langle B_\phi\rangle = 0$ isocontour.}
\label{fig:taylerNAME}
\end{figure}

Figure \ref{fig:taylerNAME} shows a time-latitude representation of the $m=1$ stability criterion (eq.~\ref{eq:new_m1}), constructed at the upper extent of the stable layer in the simulation ($r/R=0.696$, blue line on Figures \ref{fig:structure}, \ref{fig:zonesA} and \ref{fig:zonesB}) over the same 200$\,$yr subinterval as Figure \ref{fig:structure1}D. Areas in gray are stable, and the red-to-yellow color scale encodes the magnitude of the $m=1$ Tayler stability criterion, i.e., the LHS of eq.~(\ref{eq:new_m1}). The superimposed thick black lines are the NAME time series for the Northern and Southern hemispheres, the latter assigned negative values for the purpose of clarity and symmetry. The thin dashed line is the $\langle B_\phi\rangle=0$ isocontour, and are useful in identifying the spatiotemporal unfolding of magnetic polarity inversion at this depth.

The large-scale axisymmetric toroidal field does turn out to be stable at most latitudes and phases of the cycle, the stability criterion being violated mostly in the vicinity of the polarity inversion line, and at high latitudes. As argued earlier, marginal stability is in fact expected across most of the domain in our nonlinearly-saturated, statistically stationary simulation. One may however also expect the linear stability criteria to be violated wherever and whenever the instability is turning on, and has not yet had time to reach saturation and alter the background magnetic field profile. Taken at face value, Figure \ref{fig:taylerNAME} indicates that the Tayler instability is first triggered at polar latitude, which coincides temporally with the onset of the growth phase in the NAME proxy (thick black lines), and subsequently moves to progressively lower latitudes. Once the ``instability front'' reaches around $\pm 60^\circ$ latitude, corresponding to the poleward edge of the mid-latitudes magnetic field bands, the front becomes more slanted, suggesting that the higher magnetic energy available in the field bands accelerates the development of the instability, as evidenced by the crossing horizontal and vertical thin black lines on Figure \ref{fig:taylerNAME}. This is also when the NAME proxy reaches its peak before beginning to drop, signaling that the (non-axisymmetric) instability is now turning itself off via the destruction of the axisymmetric toroidal field bands making up its energy reservoir. After the instability shuts off, there follows a ``quiet'' interval during which the axisymmetric toroidal magnetic bands start to rebuild once again, up to the point where the instability will once again be triggered at high latitudes.

Interestingly, a plot similar to Figure \ref{fig:taylerNAME} but constructed for the $m=0$  stability criterion (\ref{eq:new_m0}) reveals a  spatiotemporal pattern closely resembling the $m=1$ case, but with reduced amplitude, consistent with the suppression, by rotation, of the axisymmetric Tayler instability in favor of its non-axisymmetric $m=1$ cousin, as suggested by the analysis of \citet{1985MNRAS.216..139P}. 

\section{Additional numerical experiments and a plausible scenario}
\label{sec:scenario}

In this paper we have investigated the possible occurrence of non-axisymmetric MHD instabilities developing in the subadiabatic, stably stratified fluid layers underlying the convectively unstable layer in the EULAG-MHD ``millenium simulation'' described in \citep{Passos:2014kx}. 
The cyclic, phase-lagged waxing and waning of the magnetic energies associated with the axisymmetric toroidal and non-axisymmetric total magnetic field components, as extracted from the stable layer in the simulation, bears a striking resemblance to that characterizing the magnetoshear instability studied by \citet{Miesch:2007fp} 
using a shallow-water MHD model with forced differential rotation and poloidal magnetic component. In our simulation this forcing occurs naturally through Reynolds stresses and turbulent dynamo action taking place in the overlying convecting layers. 
Our analysis 
suggests that something akin to this magnetoshear instability is operating in our simulation. \ch{The analysis of energy transfers unambiguously shows that the magnetic $m=1$ mode grows by receiving energy from the latitudinal differential rotation, and that the dominant magnetic $m=3$ mode mainly feeds from the radial structure of the large-scale axisymmetric magnetic field.}

Motivated 
by the rather weak latitudinal differential rotation characterizing the stable layers in our simulation, we also investigated the possibility that the Tayler instability be operating \ch{along with} 
the magnetoshear instability. Further motivation is also found in some recent numerical studies \citep{Brun:2006bb,Rogers:2011cx,2007ApJ...667L.207P} 
which have uncovered various elements of evidence for the action of this instability in a similar solar context. The Tayler instability does not require differential rotation, as it taps only into the magnetic energy of the underlying large-scale magnetic field to power itself, and is particularly prone to develop in the vicinity of the magnetic symmetry axis. Using linear stability as a guide, we have found some suggestive evidence for the action of the Tayler instability as an agent contributing to the destruction of the large-scale toroidal magnetic field bands building up in the simulation, especially on their poleward edge.

Whatever the exact nature of the instability that may be at play in the stable layers of our simulation, the crucial question is: does it play a significant ---or maybe even important--- role in the global dynamo process leading to amplification and polarity reversals of the large-scale magnetic field? The fact that the Poynting flux remains downward directed at all phases of the magnetic cycle and at all latitudes in the lower convection zone and overshoot layer certainly suggests that there is no direct upward electromagnetic feedback from the stable layer into the convecting fluid layers in the simulation. The obvious empirical test, running a simulation without the stably stratified fluid layer but otherwise identical, is unfortunately inconclusive. Such a simulation does \emph{not} produce a decadal large-scale cycle, but the differential rotation and cyclonic character of turbulence in the bottom of the convection zone also turn out markedly different, due to the boundary conditions that must be imposed there. It is therefore impossible to tell whether the lack of large-scale magnetic cycle is due to a different mode of dynamo action within the convection zone, rather than to the absence of the stable layer. \ch{\citet{2001ApJ...551..536D} also proposed that the tachocline instabilities could generate a net kinetic helicity leading to a local $\alpha$-effect contributing in situ to the build-up of the local poloidal field. In our simulation the kinetic helicity pattern associated with the instability is compatible with the \citet{2001ApJ...551..536D} scenario, but the estimated amplitude of the associated  $\alpha$-effect, computed following the methodology introduced by these authors, remains far too small to have any influence on the magnetic cycle timescales.}

\begin{figure}[hp]
\includegraphics[width=\linewidth]{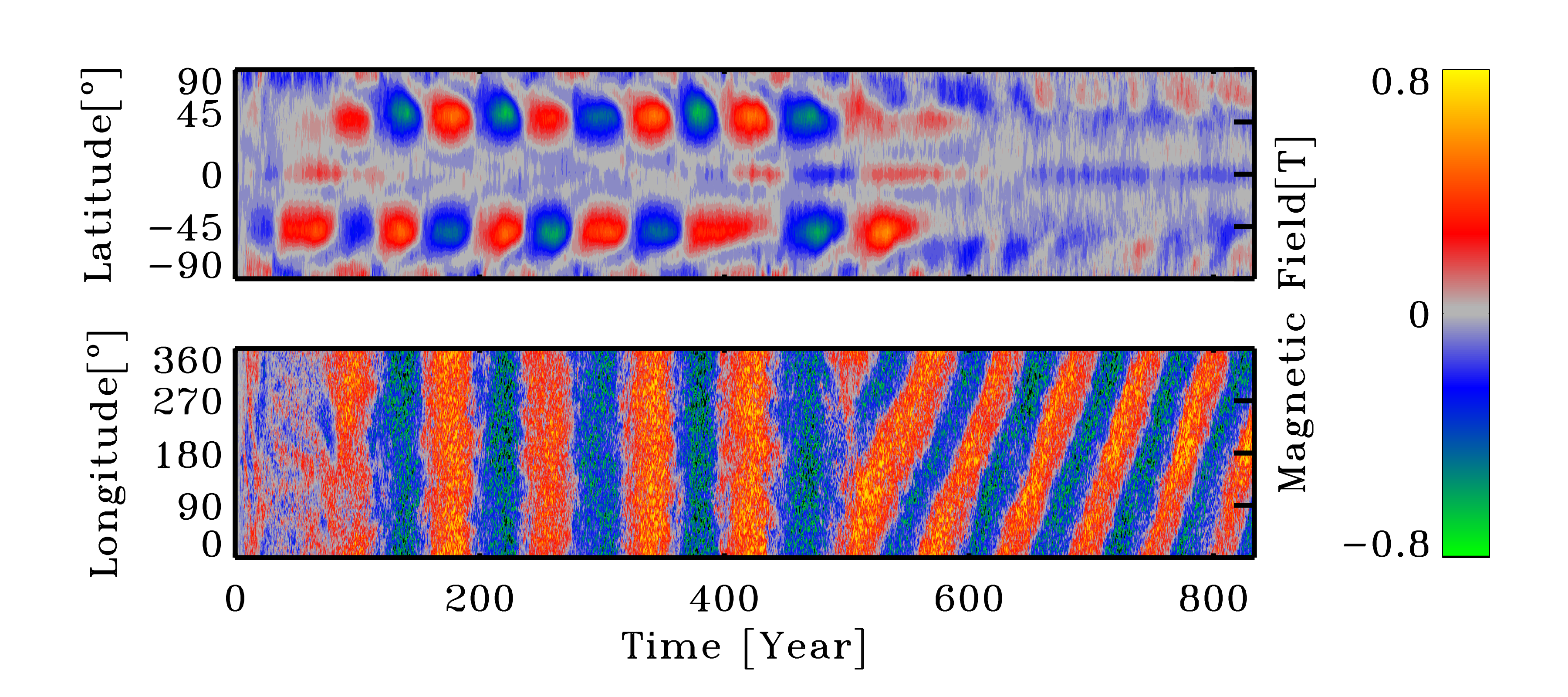}
\caption[Global wave number change]
{Time-latitude diagram of the zonally-averaged toroidal magnetic field at $r/R=0.711$ in a EULAG-MHD simulation identical to the millenium simulation, except for a more strongly subadiabatic polytropic profile in the stable layer. The bottom panel shows a time-longitude diagram extracted at $r/R=0.711$ and $45^\circ$ latitude North, in the same simulation. The apparent disappearance of the cycle on the top panel is a consequence of the dynamo switching to a non-axisymmetric large-scale mode, still undergoing polarity reversals with respect to its own
symmetry axis (see text).}
\label{fig:mhd70}
\end{figure}

Another option for numerical experimentation is to retain the stable layer, but alter its degree of subadiabatic stratification. Figure  \ref{fig:mhd70} shows the result of such an experiment. This simulation is in all aspects identical to the millenium simulation used herein, except that now the polytropic index varies linearly with depth from a value 1.5 at the base of the convecting layers up to 3.0 at the base of the domain, whereas in the millenium simulation runs the corresponding linear variation is from 1.5 to 2.5. \cch{As a result, this simulation is more stably stratified everywhere in the radiative zone compared to the millenium simulation.} The top panel of Figure \ref{fig:mhd70} shows a time-latitude diagram of the zonally-averaged toroidal magnetic field at $r/R=0.711$. Starting again from a low amplitude random initial condition, the simulation develops a large-scale magnetic cycle that is essentially identical to that characterizing the millenium simulation, showing the same cycle period, magnetic amplitude, and equatorial antisymmetry. However, after 400$\,$yr of simulation time, the Southern hemisphere fails to reverse, but the cycle picks up again and is now apparently showing symmetry about the equator. This is short lasting, as by $\simeq 500\,$yr the Northern hemisphere shuts down, followed by the Southern at $\simeq 550\,$yr.

Has the simulation fallen here into a Maunder Minimum-like state of strongly suppressed cyclic activity? The answer is no, as evidenced on the bottom panel of Figure \ref{fig:mhd70}. This shows a time-\emph{longitude} diagram of the toroidal magnetic component, \emph{not} averaged zonally, and extracted at $45^\circ$ latitude and the same depth as the top panel. Far from vanishing, the large-scale magnetic cycle persists with similar amplitude and a shorter half-period of $\simeq 30\,$yr, but now its symmetry axis has undergone a large tilt with respect to the rotation axis. 

At the very least, these numerical experiments indicate that the structure of the stable layers affects the long term stability of the various types of large-scale cyclic dynamo modes that can materialize in these EULAG-MHD simulations. But how about the polarity reversal taking place in the more solar-like axisymmetric mode of large-scale dynamo action arising in the millenium simulation? The following ``global dynamo scenario'' is consistent with all analyses presented in this paper, with the global 3D MHD simulations of \citet{2013ApJ...778...11M}, 
as well as with the more geometrical restrictive simulations of \citet{Miesch:2007fp} 
and \citet{Rogers:2011cx}: 
\begin{enumerate}
\item Dynamo action is driven primarily within the convection zone, through the differential rotation and turbulent electromotive force materializing therein. This is consistent with the analyses of \citet{Racine:2011gh} and \citet{2013ApJ...768...16S}, 
which indicate a mode of large-scale dynamo action resembling the so-called $\alpha^2\Omega$ dynamos of mean-field theory;
\item Downward pumping of the magnetic field produced in the convecting layers leads to the buildup of strong zonally-aligned magnetic field bands in the overshoot layer and upper reaches of the underlying stably stratified fluid layer, where further amplification of the toroidal magnetic component takes place through shearing of the poloidal magnetic field by differential rotation; downward pumping is observed in virtually all extant MHD simulations of solar convection, and differential rotation shear is a key ``ingredient'' of most extant solar cycle models;
\item Once the toroidal magnetic field bands in the stable layers reach sufficient strength, likely of the order of a few tenths of Tesla, MHD instabilities set in, destabilizing the magnetic field bands and accelerating their dissipation;
\item Meanwhile the global magnetic polarity has reversed in the convecting layers, and downwards pumping of magnetic field of opposite polarity to that having formerly built up in the stable layer begins, eventually leading to polarity reversals therein as well, closing the dynamo loop towards step 1 above.
\end{enumerate}

Additional simulations and analyses are underway to further validate this scenario, which at this juncture remains speculative but plausible. From the point of view of kinematic mean-field and mean-field-like dynamo models of the solar cycle, the instabilities for which we have presented evidence in this paper can be considered as contributing to the enhanced turbulent magnetic diffusivity that is essential for such dynamo models to produce solar-like cycles with decadal periods.

\medskip\noindent {\bf Acknowledgements}

We wish to thank Sacha Brun useful discussions on the Tayler instability and
insightful comments upon reading a first draft of this paper. \ch{We also thank an anonymous referee for very useful comments and suggestions that helped strengthen the analysis developed in this work.}
The EULAG-MHD ``millenium simulation'' was originally designed and executed by Mihai Ghizaru. N. Lawson wishes to thanks the physics department of Universit\'e de Montr\'eal for the award of a graduate fellowship. A. Strugarek is a National Postdoctoral Fellow of
the Canadian Institute for Theoretical Astrophysics. P. Charbonneau is supported primarily by a Discovery Grant from the Natural Sciences and Engineering Research Council of Canada. 
All simulations reported upon in this paper were performed on the computing infrastructures of Calcul Qu\'ebec, a member of the Compute Canada consortium.

\bibliographystyle{yahapj}
\bibliography{/Users/astrugar/WORK/MyPapers/mybib}

\end{document}